\documentclass[10pt,english,pre,nofootinbib,reprint, groupedaddress,showpacs,showkeys,preprintnumbers,floatfix]{revtex4-1}

\usepackage[latin9]{inputenc}
\usepackage[a4paper]{geometry}
\usepackage{xcolor}
\usepackage{pdfcolmk}
\usepackage{graphicx}
\usepackage{babel}
\usepackage{mathrsfs}
\usepackage{amsmath}
\usepackage{amssymb}
\usepackage{esint}
\usepackage[toc,page]{appendix}

\PassOptionsToPackage{normalem}{ulem}
\usepackage{ulem}
\usepackage{dsfont} 
\usepackage[unicode=true,pdfusetitle,
 bookmarks=true,bookmarksnumbered=false,bookmarksopen=false,
 breaklinks=false,pdfborder={0 0 1},backref=false,colorlinks=false]
 {hyperref}
\usepackage{tikz}

\makeatletter
\newcommand{\Rmnum}[1]{\expandafter\@slowromancap\romannumeral #1@}

\usepackage{bbold}

\let\textquotedbl="

\makeatother

\begin{document}

\title{Inference of signals with unknown correlation structure from nonlinear measurements}

\author{Jakob Knollm\"uller, Theo Steininger and Torsten A. En\ss lin}

\affiliation{{\small{}Max-Planck-Institut f\"ur Astrophysik, Karl-Schwarzschildstr.~1,
85748 Garching, Germany}\\
Ludwig-Maximilians-Universit\"at M\"unchen, Geschwister-Scholl-Platz{\small{}~}1,
80539 Munich, Germany}

\begin{abstract}
We present a method to reconstruct autocorrelated signals together with their autocorrelation structure from nonlinear, noisy measurements for arbitrary monotonous nonlinear instrument response. In the presented formulation the algorithm provides a significant speedup compared to prior implementations, allowing for a wider range of application. The nonlinearity can be used to model instrument characteristics or to enforce properties on the underlying signal, such as positivity. Uncertainties on any posterior quantities can be provided due to independent samples from an approximate posterior distribution. We demonstrate the methods applicability via simulated and real measurements, using different measurement instruments, nonlinearities and dimensionality.
\end{abstract}

\maketitle
 
\section{Introduction}
The challenge of reconstructing signals from noisy measurements is omnipresent in science and technology. A large variety ranging from general to highly specific methods are applied to recover the concealed quantities from the data. When having an excellent signal-to-noise ratio, it might be sufficient to back project the data using a description of the instrument and perform a deconvolution with the instrument kernel. This procedure is called the filtered back projection and is widely used in modern medical imaging applications \citep{filteredbackprojection}. The better the understanding of the data in form of prior knowledge about the process of data generation, the better the signal's reconstruction. One noteworthy example of this is the large topic of compressed sensing \citep{compressed, robustCS}. It can fully recover sparse signals, given some further properties, in only partially sampled data, which is possible due to the Nyquist-Shannon sampling theorem \citep{nyquisttelegraph, shannoncommunication}.  In many situations, however, one does not want to assume sparsity or high signal-to-noise ratio. Instead, it is necessary to use general models, capturing only justifiable features of the signal to avoid a biased reconstruction.

 One such general property is autocorrelation. Many signals we observe are generated by processes following temporal or spatial correlation. If the correlation structure is known and exploited, it significantly increases the fidelity of the reconstruction. If it is the only known property of the signal, in the case of linear measurement and Gaussian noise, one ends up with the well-known Wiener filter \citep{Wiener} due to maximum entropy considerations. As it builds the foundation of our considerations a detailed discussion of the Wiener Filter can be found in Appendix \ \ref{sec:WienerFilter}, also to introduce the information field theory (IFT) formalism \citep{IFT} used in this paper. 

Unfortunately, by far not for all processes the correlation structure is known. In many cases it is even the quantity of interest itself, as it provides deep insights into the underlying process generating the signal. This can, for example, be done via Exact Bayesian Covariance Estimation and Signal Reconstruction for Gaussian Random Fields (MAGIC)  \citep{MAGIC} which estimates the exact posterior using Gibbs sampling. However, this method requires huge computational resources, which limits it to only a small selection of problems where such precision is necessary. An approximate, variational approach is the critical filter \citep{ensslinfrommert}, which as well reconstructs the signal and its correlation structure, using homogeneity and isotropy arguments. Its concepts and technical details are outlined in Appendix \ \ref{sec:CriticalFilter}. The fundamental assumption underlying these approaches is the existence of statistical homogeneous  autocorrelation, without specifying it any further. It has been successfully applied in various forms in the context of astrophysical imaging \citep{D3PO, RESOLVE}.
 
So far the critical filter, due to tremendous computational costs, was limited in its applicability to relatively small problems, despite being significantly faster than methods based on Gibbs sampling. In this paper we will propose a mathematical reformulation of its internal signal representation to overcome such limitations and drastically speed up algorithmic convergence, which will be outlined in \ref{sec:reformulation}. With this we hope to open it for a large variety of possible applications. This is achieved by changing from an indirect reconstruction scheme to direct communication of the correlation parameters with the data. In addition we will extend the model to arbitrary nonlinear measurements. In this situation no analytic solution to the inference problem is available, which is why we will rely on variational inference schemes, minimizing the Kullback-Leibler (KL) divergence \citep{KLdivergence} between the true posterior distribution and its approximation. The KL divergence itself is also not calculable for arbitrary nonlinear response functions. Instead, we minimize a statistical estimate of it, using samples from the approximate posterior, from which we can sample independently. Those samples can also be used to provide uncertainty estimates to arbitrary posterior quantities, such as the local reconstruction fidelity.

 This turns the entire algorithm into a stochastic minimization problem. Restricting oneself to only monotonous response functions, unique solutions can be obtained. The flexibility of the presented approach with respect to response functions can be used to model nonlinear, instrument specific characteristics or to enforce additional signal characteristics, such as positivity or strong spatial variability.
 
After a brief summary of the the required algorithmic steps we present its application to two distinct examples.
As the model is formulated in the abstract language of information field theory (IFT) \citep{IFT}, it can be straightforwardly applied in a large variety of contexts, independent of the number of spacial dimension or measurement instruments.  The first example demonstrates the method's capability of dealing with rather extreme nonlinearities, containing jumps and flat regions. In the second example, a real data application, we reconstruct the supernova remnant 3C391 from VLA radio interferometric data, and additionally perform a noise calibration during inference.

\section{Nonlinear Measurements}
The situation we want to discuss in this paper is the nonlinear measurement of an autocorrelated  Gaussian signal $s$ with additive Gaussian noise $n$. The nonlinearity is described by a monotonous, local function $f(s)$, which modifies each value of $s$. The nonlinear function can serve two distinct purposes. On the one hand we can use it to model nonlinear instrument characteristics, such as detector saturation. On the other hand we can use it to express expected non-Gaussian signal characteristics. We can, for example, use it to enforce positivity or allow for exponential variations between locations using the exponential function. The latter corresponds to a lognormal model, which is widely used in astrophysical image reconstructions. The measurement itself is characterized by the linear measurement operator $R$, encoding a detailed description of the measurement process.  The data $d$ is than related to the field $s$ via the measurement equation
\begin{align}
\label{eq:dataequation}
d = R f(s) + n \text{.}
\end{align}
Here $f$ is applied ti each field value $s_x$ separately, such that $f(s)_x = f(s_x)$. $R$ is the linear (part of) the instrument response, which maps the continuous field $f(s)$ into a discrete data vector.
Assuming the noise covariance is known, we can marginalize out the noise, which provides us with a Gaussian likelihood. We will encode the autocorrelation of the signal field $s$ in a Gaussian signal prior with correlation structure $S$. In order to infer the signal field $s$ given the data, we combine the likelihood with the prior according to Bayes theorem. More conveniently we simply add up the log-likelihood and the log-prior. It reads
\begin{align}
\mathcal{H}(d, s) \:\widehat{=} &\: \frac{1}{2} \left(d-Rf(s)\right)^\dagger N^{-1}\left(d-Rf(s)\right)\nonumber \\
&+ \frac{1}{2} s^\dagger S^{-1} s \text{.}
\end{align}
The first line corresponds to the Gaussian log-likelihood, the second term to the Gaussian log-prior. Here the $\widehat{=}$ indicates that signal independent constants are dropped, as they vanish during the normalization. The expression above corresponds to the un-normalized log-posterior, or in the IFT language the problem Hamiltonian, as it corresponds to an energy functional for fields.
For a linear function $f(s)$ this is the Wiener Filter problem, which is analytically tractable. A detailed derivation, starting with the linear data equation and containing all relevant steps can be found in Appendix \ref{sec:WienerFilter}.

The more general problem described above this is the nonlinear Wiener Filter problem. Here the normalization is not feasible for arbitrary functions. The usual approach would then be an approximation, which maximizes the Hamiltonian above to obtain the Maximum Posterior (MAP) estimate.

\section{Critical filtering}
We want to extend this model to a priori unknown correlation structures $S$ of $s$, using the model of the critical filter\ \citep{smoothpower}, which has already been successfully applied in multiple astrophysical imaging problems \citep{D3PO, RESOLVE}. It assumes a priori homogeneity and isotropy, which allows the correlation structure to be expressed as a power spectrum in the harmonic basis, i.e. the Fourier space. This power spectrum is discretized into individual bins, of which the coefficients are inferred. The covariance is expressed in terms of a logarithmic power spectrum to ensure positive definiteness of the correlation structure. A prior assumption to the spectrum is smoothness on a logarithmic scale in Fourier space, which is expressed in terms of the $L_2$ norm of the second derivative. Combining this model with the nonlinear Wiener filter model from the previous section, we obtain the full critical filter Hamiltonian for nonlinear measurements. It reads:
\begin{align}
\label{eq:nlcf_ham}
\mathcal{H}(s,\tau\vert d) \widehat{=}& \frac{1}{2}(d - Rf(s))^\dagger N^{-1} (d-Rf(s)) \nonumber \\ 
 &-  \frac{1}{2} \varrho^\dagger \tau + \frac{1}{2} s^\dagger  \mathbb{F}^\dagger\left( \widehat{\mathbb{P}^\dagger e^{-\tau}}\right)  \mathbb{F}s \nonumber \\
 & + \frac{1}{2 \sigma^2} \tau^\dagger \Delta^\dagger \Delta \tau
\end{align}
The first line originates from the likelihood, the second line corresponds to the prior of $s$, now depending on the logarithmic power spectrum $\tau$. The $\mathbb{F}$ operator indicates Fourier transformation and the isotropic projection $\mathbb{P}$ relates the power spectrum to the correlation structure to the diagonal of $S$ in the harmonic domain. The $\widehat{ \: \:}$- symbol raises the diagonal to a diagonal operator. The third line implements the smoothness prior via the second derivative $\Delta$ on a logarithmic scale and expected deviation $\sigma$ from it. A detailed derivation containing various subtleties of practical relevance is outlined in Appendix \ref{sec:CriticalFilter}.

In order to achieve convergence, the maximum posterior approach is insufficient, instead one has to perform a variational inference, capturing uncertainties within the posterior signal field $s$. 
The inference problem itself is then solved by iterative optimization with respect to the signal field $s$ and logarithmic power spectrum $\tau$. This procedure itself is dreadfully slow due to the strong interdependence between the correlation structure and the corresponding signal field. In one step the signal is inferred for the current power spectrum, in the next step this signal field is used to update the correlation structure. In each step only tiny improvements are possible, especially in the regime of high noise, where the data is inconclusive. This problem can be illustrated as high dimensional, diagonal valleys in the energy landscape, in which we are looking for the minimum. The alternating procedure leads to a zigzag movement following only slowly the diagonal valley downhill.

To overcome this problem we will propose a reformulation of the model described above, which strongly decouples the inference problem of the signal field and its correlation structure. 
\section{The new Critical Filter}
\label{sec:reformulation}

According to our prior distribution the signal $s$ is drawn from a Gaussian distribution, whose covariance matrix is diagonal in the harmonic basis. We can draw a sample from a Gaussian distribution by first drawing  a point-wise independent, Gaussian, white excitation field $\xi$ with unit variance $\mathbb{1}$ in its eigenbasis. Weighting this excitation field with the square root of its eigenvalues provides a sample from the initial Gaussian distribution. The square root of the eigenvalues correspond to the standard deviation of each individual mode, serving as the amplitude of the excitation. Introducing the amplitude operator $A$ and the logarithmic amplitude spectrum $\alpha$ will simplify the notation. They are defined as
 \begin{align}
 \label{eq:amplitude}
 A \equiv \mathbb{F}^\dagger\widehat{\mathbb{P}^\dagger e^{\alpha}}  &\equiv \mathbb{F}^\dagger\widehat{\mathbb{P}^\dagger e^{\frac{1}{2} \tau}}
 \text{, and}\\
 A A^\dagger & = S 
 \end{align}
Therefore, the signal field can be rewritten as 
 \begin{align}
 s &=A \xi \text{.}
 \end{align}
This has two important consequences for the formulation of our problem. The first one concerns the signal field prior, which becomes
\begin{align}
\frac{1}{2}s^\dagger S^{-1} s &\rightarrow \frac{1}{2}\xi ^\dagger A^\dagger S^{-1} A \xi = \frac{1}{2}\xi ^\dagger \mathbb{1} \xi \text{,}
\end{align}
using the identity in Eq.\ \ref{eq:amplitude}. This expression does not depend on the power spectrum any longer and corresponds to a $L_2$ regularization of the excitation. The change of variables also removes the prior normalization, which also contains the power spectrum due to the functional determinant. 

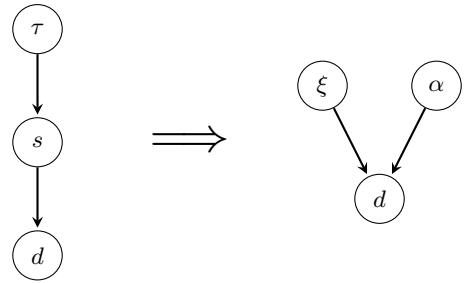
\begin{figure}[h]
    \centering
        \begin{tikzpicture}
            [c/.style={circle,minimum size=2em,text centered,thin},
             r/.style={rectangle,minimum size=2em,text centered,thin},
             v/.style={->,shorten >=1pt,>=stealth,thick}, 
             arrow/.style={-latex, shorten >=1ex, shorten <=1ex, bend angle=45}]
            \node(b)at(-3.25,5)[c,draw]{$\tau$};
            \node(c)at(-3.25,3.5)[c,draw]{$s$};
            \node(d)at(-3.25,2)[c, draw]{$d$};
            \node(e)at(2,4.25)[c,draw]{$\alpha$};
            \node(f)at(0.5,4.25)[c,draw]{$\xi$};
            \node(g)at(1.25,2.75)[c, draw]{$d$};
            \node(h)at(-1.25,3.5)[scale=2]{$\Longrightarrow$};
	    \draw[v](b.south)--(c);
	    \draw[v](c.south)--(d);
	    \draw[v](e)--(g);
	    \draw[v](f)--(g);	    	  
        \end{tikzpicture}
        \flushleft
        \caption{Causality structure of the classical critical filter (left) and of its reformulation (right).}
   \label{fig:diagram} 
\end{figure}

Replacing the signal field in the likelihood, too, and using $\alpha = \frac{1}{2} \tau$ we can rewrite the full problem Hamiltonian as
\begin{align}
\label{eq:hamiltonian}
\mathcal{H}(\xi,\alpha\vert d) =& \frac{1}{2}(d-Rf( A\xi))^\dagger N^{-1}(d-Rf(A\xi))\nonumber \\
&+ \frac{1}{2} \xi^\dagger \mathbb{1} \xi + \frac{2}{ \sigma^2} \alpha^\dagger \Delta^\dagger \Delta \alpha \text{.}
\end{align}
The old formulation reconstructs in Eq.  \ref{eq:nlcf_ham} the signal, which just is the product of excitation and the square root of the power spectrum, and uses it to estimate the power spectrum. This shows the high coupling between those two quantities. In this new formulation the excitation and amplitudes are two far more independent quantities, as the excitation is a priori white noise, and the logarithmic amplitude spectrum is smooth. Both quantities now also appear in the likelihood, which allows them to talk to the data directly during the inference, compared to previously, where the power spectrum was only indirectly inferred. Fig. \ref{fig:diagram} illustrates the parameter dependence in the previous indirect and the new, direct model. 

\begin{figure}
\includegraphics[scale=0.57]{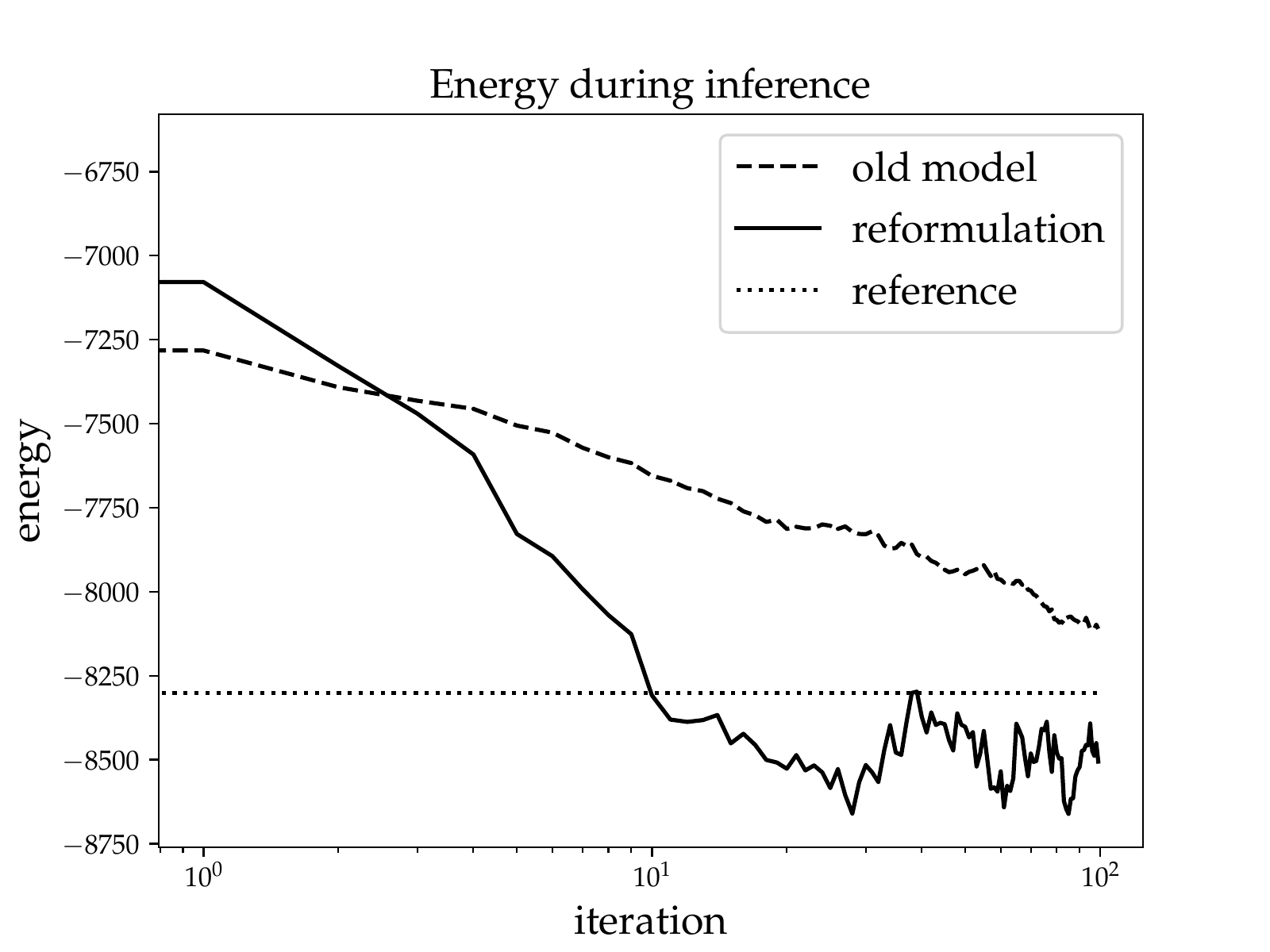}
\caption{The convergence behavior in terms of the KL-divergence for the old model and the reformulated version as a function of iteration for $100$ steps. The reference KL uses the Wiener filter solution for the map and the correct power spectrum.}
\label{fig:convergence}
\end{figure}

The effect of this change on the numerical convergence of the resulting algorithm is drastic and can be seen in Fig.\ (\ref{fig:convergence}). The same setup and initial conditions for the linear measurement of a one dimensional signal with resolution $1024$ is used for both cases.
For the inference we will optimize a stochastic estimate of the Kullback-Leibler divergence between the true posterior distribution and a simpler, approximate posterior distribution to fit its parameters. A detailed discussion of all quantities involved will follow in the next section, at this place we only want to give the motivation for the reformulation.
Fig. \ref{fig:convergence} shows that both methods decay roughly as power laws with two clearly distinct slopes in the energy decrease per decade of iterations. The new method  already shows signs of convergence after twenty iterations, whereas the old reconstruction takes significantly longer. Even after one hundred iterations it is far from convergence. Linear extrapolation in Fig. \ref{fig:convergence} roughly hints that the direct method requires two orders of magnitude less iterations in this specific scenario. It is worth to point out that the computational effort for one individual step for both methods is comparable.

After the new method has converged, the stochasticity of the target functional, which estimates the KL, starts to show up. The further evolution of the algorithm is dominated by random fluctuations due to this stochasticity. Using more samples to estimate the KL would reduce this behavior, but it would also increase the computational cost. The reference energy shown in Fig \ref{fig:convergence} corresponds to the KL of the Wiener filter solution using the correct power spectrum which was used to generate the data. It is not surprising that our reformulated method slightly undercuts this energy, as there might be better fitting correlation structures given the concrete realization.
In the following we will discuss the derivation of the inference algorithm using the reformulated critical filter model for nonlinear measurements as defined in Eq.\ \ref{eq:hamiltonian}.

\section{Approximating the posterior}
We are not able to calculate the posterior of our parameters given the data analytically. Hence, we have to think about an approximation to the true posterior distribution. As for the classical critical filter approach, using only the MAP estimate turned out to be insufficient. Furthermore, we have to take the uncertainty of the reconstructed excitation field into account. The simplest way to do this is to approximate the posterior distribution by the product of a Gaussian distribution for the excitation and a point estimate for the amplitude spectrum. We can write this as
\begin{align}
\widetilde{\mathcal{P}}(\xi, \alpha\vert d) = \mathcal{G}(\xi-t,\Xi)\delta(\alpha-\alpha_*) \text{.}
\end{align}
The usual approach now would be to minimize the Kullback-Leibler divergence  \citep{KLdivergence} between this approximative posterior and the true posterior from  Eq.\ \ref{eq:hamiltonian}, which lets us avoid calculating the posterior normalization as is analytically intractable. It is defined as
\begin{align}
\label{eq:KL}
\mathcal{D}_\mathrm{KL}&\left[\widetilde{\mathcal{P}}(\xi,\alpha\vert d)\vert\vert\mathcal{P}(\xi,\alpha\vert d)\right] \equiv \nonumber \\
\equiv &\int \mathcal{D}\xi \:\mathcal{D}\alpha\: \widetilde{\mathcal{P}}(\xi,\alpha\vert d) \:\mathrm{ln}\left[\frac{\widetilde{\mathcal{P}}(\xi,\alpha\vert d)}{\mathcal{P}(\xi,\alpha\vert d)}\right] \text{.}
\end{align}
However, we will not use this to estimate all parameters of our approximate posterior as it still contains numerically problematic objects. We will pursue a hybrid approach, estimating the parameters of the Gaussian using the Laplace approximation by maximizing the Hamiltonian with respect to the excitation. This maximum is then used as the approximate posterior excitation mean $t$. The gradient of the Hamiltonian with respect to $\xi$ reads
\begin{align}
\label{eq:ex_grad}
\frac{\delta \mathcal{H}(\xi\vert d)}{\delta \xi} =& -(d-Rf( A\xi))^\dagger N^{-1}R f'(A\xi)A\nonumber \\
 &+ \mathbb{1} \xi \text{,}
\end{align}
with $f'(s) = \frac{\delta f(s)}{\delta s}$ being the derivative of the function $f$ with respect to the signal field, evaluated at the current position. This derivative represents a linear operator. In the Laplace approximation the inverse curvature at the maximum of the Hamiltonian is used to approximate the correlation structure. Deriving the expression above by the excitation for a second time yields
\begin{align}
\label{eq:ex_curv}
\frac{\delta^2 \mathcal{H}(\xi\vert d)}{\delta \xi \delta \xi^\dagger} =& -(d-Rf( A\xi))^\dagger N^{-1}R f''(A\xi)AA\nonumber \\
&+  A^\dagger f'(A\xi)^\dagger R^\dagger N^{-1}R f'(A\xi) A \nonumber \\
&+\mathbb{1} \text{.}
\end{align}
The first term, containing the second derivative $f''$ is problematic, as it is not necessarily positive definite and can introduce negative curvature, which is prohibited for a covariance operator. However, we expect its contribution to be small in the vicinity of the minimum. It contains the residual $d-Rf(A\xi)$, which measures how compatible the current reconstruction of the signal field $ A\xi$ is to the data. This quantity is minimized  during inference under model constraints. Because of this we drop this term. For functions with extremely large second derivatives in the relevant parameter range this might not be justified, but most reasonable functions should not be affected by this. 
Once we obtained $t$ by maximizing the Hamiltonian, we evaluate the modified curvature at this position as an estimate of the posterior correlation structure.
\begin{align}
\label{eq:approx_WF_cov}
\Xi^{-1} =A^\dagger f'(At) R^\dagger N^{-1}R f'(At) A
+\mathbb{1}
\end{align}
So far we have an approximation for the excitation's posterior distribution given an amplitude spectrum. We want to use this to infer the amplitude spectrum itself, correcting for uncertainties in the excitation. We can use the KL divergence to infer the most reasonable point estimate $\alpha_*$. 
Note that this is not equivalent to a joint MAP estimate of both quantities, although we used MAP to calculate the approximate mean posterior excitation. The integration over $\alpha$ in Eq.\ \ref{eq:KL} just inserts $\alpha_*$ as we have to integrate over the delta distribution. What remains is the Gaussian expectation value over the problem Hamiltonian and the Hamiltonian of the Gaussian itself, which is an entropy.
\begin{align}
\mathcal{D}_\mathrm{KL}
= &  \:\left\langle \mathcal{H}(\xi,\alpha_{*}\vert d)\right\rangle_{\mathcal{G}(\xi-t,\xi)} \nonumber  \\
&- \left\langle \mathrm{ln}\left[\mathcal{G}(\xi-t,\Xi)\right]\right\rangle_{\mathcal{G}(\xi-t,\Xi)} \text{ .}
\end{align}
The second term, the entropy, does not explicitly depend on the amplitude spectrum and will therefore not be relevant for the minimization. Hence, we drop it. What remains is the expectation value of the problem Hamiltonian over the Laplace approximated excitation distribution. The relevant KL divergence therefore reads
\begin{align}
\mathcal{D}_\mathrm{KL} \:\widehat{=}\:& \langle \frac{1}{2}(d-Rf( A_*\xi))^\dagger N^{-1}(d-Rf( A_*\xi))\nonumber\rangle_{\mathcal{G}(\xi-t,\Xi)} \\
& + \frac{2}{ \sigma^2} \alpha_*^\dagger \Delta^\dagger \Delta \alpha_* \text{.}
\end{align}
The smoothness prior has no dependence on $\xi$ and is therefore not affected by the expectation value. Only the likelihood depends on the excitation $\xi$. We have to evaluate the expectation value, however, for arbitrary functions $f(A\xi)$ this is not analytically possible. In the linear case we can calculate this expectation value, which results in the classical critical filter. For the exponential case we also have analytic formulas, but those are not tractable numerically as they involve operator exponentiation. 

An elegant solution to this for arbitrary functions is to sample the expectation value and its derivative with respect to the amplitude spectrum. As the Laplace approximation does have the mathematical structure of a Wiener filter, we do have access to independent posterior samples. We will use them to replace the full distribution with a sampling distribution of the form 
\begin{align}
\mathcal{G}(\xi - t,\Xi) \approx  \frac{1}{N}\sum_{i=1}^{N} \delta(\xi-\xi_i^*) \text{.}
\end{align}
Any expectation value under this distribution just becomes the mean of this quantity over the set of samples, transforming highly complex expressions into simple averages.
How to obtain those samples is explained in Appendix \ref{sec:sampling}. For now, we continue to calculate the gradient of the KL with respect to the amplitude spectrum, ignoring the Gaussian expectation value. The gradient reads
\begin{align}
\label{eq:amp_grad}
\frac{\delta \mathcal{D}_\mathrm{KL}}{\delta \alpha_*} = &\langle (d-Rf(A_*\xi))^\dagger N^{-1}Rf'(A_*\xi)\widehat{A_*\xi}\mathbb{P}^\dagger  \rangle_{\mathcal{G}(\xi-t,\Xi)} \nonumber \\
&+\frac{4}{\sigma^2}\Delta^\dagger\Delta\alpha_* \text{.}
\end{align}
We can use this gradient to minimize the KL divergence. To apply a Newton scheme in this minimization, we also need the curvature of the KL, which corresponds to the second derivative. Again we drop problematic terms containing second derivatives.
\begin{align}
\label{eq:amp_curv}
\frac{\delta^2 \mathcal{D}_\mathrm{KL}}{\delta \alpha_*\delta \alpha_*^\dagger} = & \langle \mathbb{P} \widehat{A_*\xi}^\dagger f'(A_*\xi))^\dagger R^\dagger N^{-1}\nonumber \\
 & Rf'(A_*\xi)\widehat{A_*\xi}\mathbb{P}^\dagger  \rangle_{\mathcal{G}(\xi-t,\Xi)} \nonumber \\
&+\frac{4}{\sigma^2}\Delta^\dagger\Delta \text{.}
\end{align}

\section{Inference Algorithm}
We have now obtained all expressions and methods to set up an iterative scheme to infer the amplitude spectrum and its excitation of a signal field from a nonlinear measurement with additive noise. We start with arbitrary parameter values for the approximate posterior excitation mean $t$ and amplitude spectrum $\alpha_*$. 

The next step is to minimize the information Hamiltonian 
\begin{align}
\mathcal{H}(\xi,\vert d, \alpha_*) \: \widehat{=} \: & \frac{1}{2}(d-Rf( A\xi))^\dagger N^{-1}(d-Rf(A\xi))\nonumber \\
&+ \frac{1}{2} \xi^\dagger \mathbb{1} \xi
\end{align}
with respect to $\xi$ to obtain an estimate for $t$ using the gradient from Eq.\ (\ref{eq:ex_grad}) and curvature (\ref{eq:ex_curv}) within a relaxed Newton scheme. In the linear case the minimum is reached after the first step as this is equivalent to the Wiener filter. For nonlinear functions one has to perform several steps, but the Newton scheme leads to fast convergence.

Once the Hamiltonian is minimized for the given amplitude spectrum $\alpha_*$, we construct the Laplace approximation by using the curvature at the minimum as approximate posterior covariance. From this distribution we calculate a set of samples $\{\xi^*\}$ using the procedure described in Appendix \ref{sec:sampling}. 
\begin{align}
\xi^* \curvearrowleft \mathcal{G}(\xi^* - t, \Xi)
\end{align}
We use this set of samples to approximate the KL-divergence by replacing the Gaussian expectation value with the sampling mean of this set.
\begin{align}
\mathcal{D}_\mathrm{KL} \:\widehat{=}\:& \langle \frac{1}{2}(d-Rf( A_*\xi))^\dagger N^{-1}(d-Rf( A_*\xi))\nonumber\rangle_{ \{\xi^*\} } \\
& + \frac{2}{ \sigma^2} \alpha_*^\dagger \Delta^\dagger \Delta \alpha_*
\end{align}
Here we can use its gradient from Eq.\ (\ref{eq:amp_grad}) and its curvature in Eq.\ (\ref{eq:amp_curv}) to minimize the divergence again using a relaxed Newton scheme. Once this minimization is converged we can adapt it and continue with another minimization of the Hamiltonian with respect to the excitation using this new amplitude spectrum, which leads to the next excitation estimate. This allows us to draw new samples, which can then be used to re-estimate the amplitudes. This is repeated until the desired convergence is reached. Once we have finished the inference we can use the approximate posterior parameters to calculate any desired quantity with uncertainties using posterior samples. Examples for such are the mean posterior signal with applied nonlinearity
\begin{align}
\langle f(s)\rangle_{\widetilde{\mathcal{P}}(s\vert d)} \approx \langle f(A_* \xi) \rangle_{\{\xi^*\}} \text{,}
\end{align} 
or the uncertainty of the correlated signal estimate
\begin{align}
\langle (s-m)^2\rangle_{\widetilde{\mathcal{P}}(s\vert d)}= \langle [A_*(\xi - t)]^2 \rangle_{\{\xi^*\}} \text{.}
\end{align}

\section{Numerical Examples} 
We demonstrate the capabilities of the derived algorithm on two different examples. The algorithm was implemented in python using the numerical information field theory package NIFTy \citep{Nifty3}, which provides the abstract structure to conveniently implement algorithms derived with IFT.

 The first example is based on one dimensional, synthetic data. It includes a rather artificial monotonous nonlinearity with piecewise changing functions, exhibiting discontinuities and flat regions, illustrated in Fig.\ \ref{fig:nonlinearity}. Using mock data, we have access to the true underlying signal, which allows us to compare it to the reconstruction. 

In the second example we reconstruct a two-dimensional image of the supernova remnant 3C391 from radio-interferometric data obtained by the VLA telescope. The nonlinearity in this case is used to model the signal. For radio sources we do want to enforce positivity and allow for spatial variations of the intensities. A typical choice for $f(s)$ to account for these requirements is the exponential function, which leads to a log-normal intensity model \citep{RESOLVE,D3PO}. Interferometric data consists of point measurements in the Fourier plane, whereby large parts of the actual image are masked. This complex response can be included straightforwardly, allowing for high quality radio reconstructions.

In the examples we will reconstruct the amplitude spectrum and excitation. However, those quantities are rather counterintuitive. For illustrative purposes we show power spectra and signal fields. The power spectrum is just the squared amplitude spectrum and the signal fields are obtained from the harmonic transformation of the amplitude weighted excitation.
\subsection{One dimensional measurement}
\begin{figure}
\begin{center}
\includegraphics[scale = 0.5]{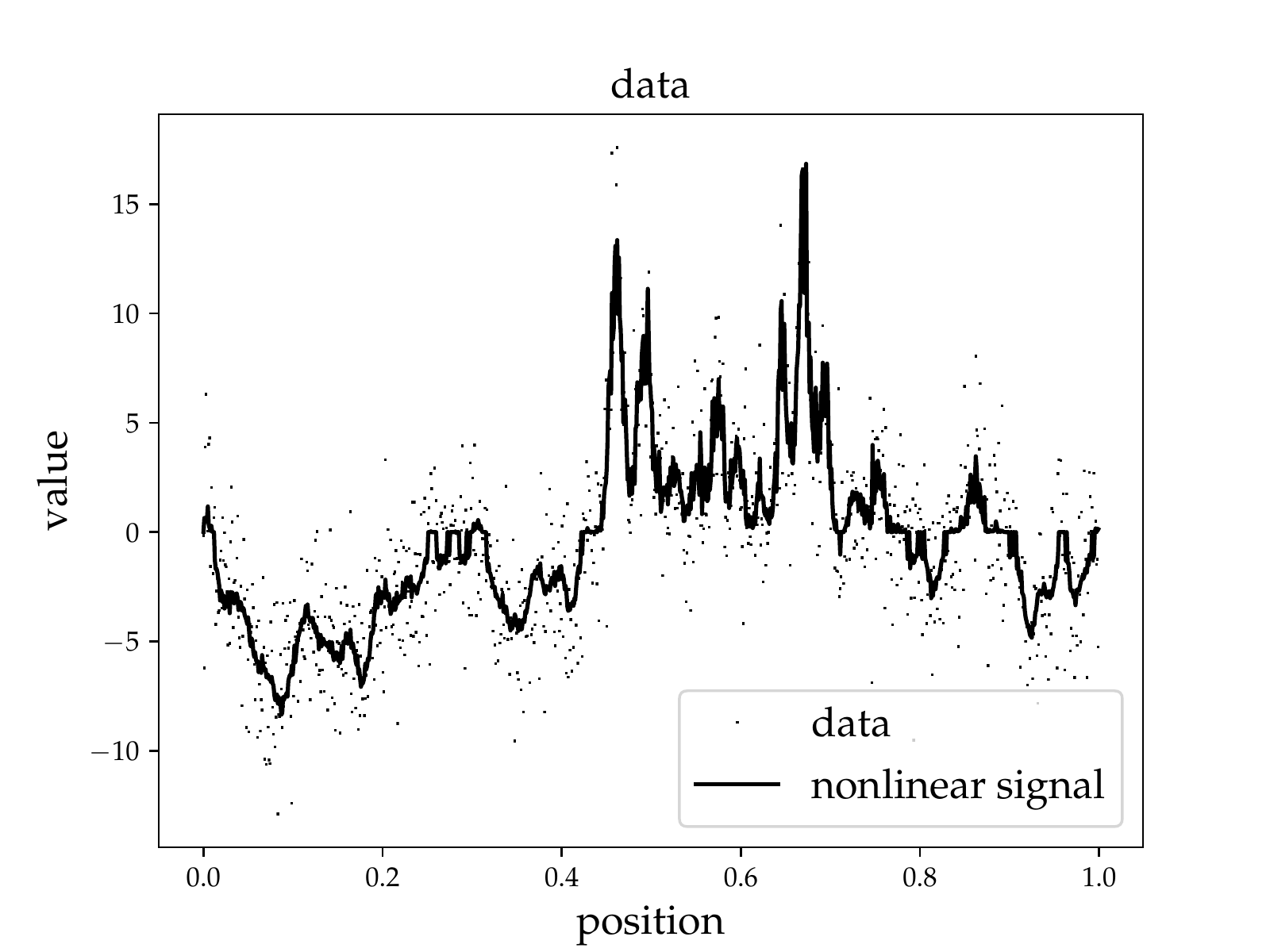}
\caption{The data and nonlinear signal for the first example. }
\label{fig:data}
\end{center}
\end{figure}

\begin{figure}

\begin{center}

\includegraphics[scale = 0.5]{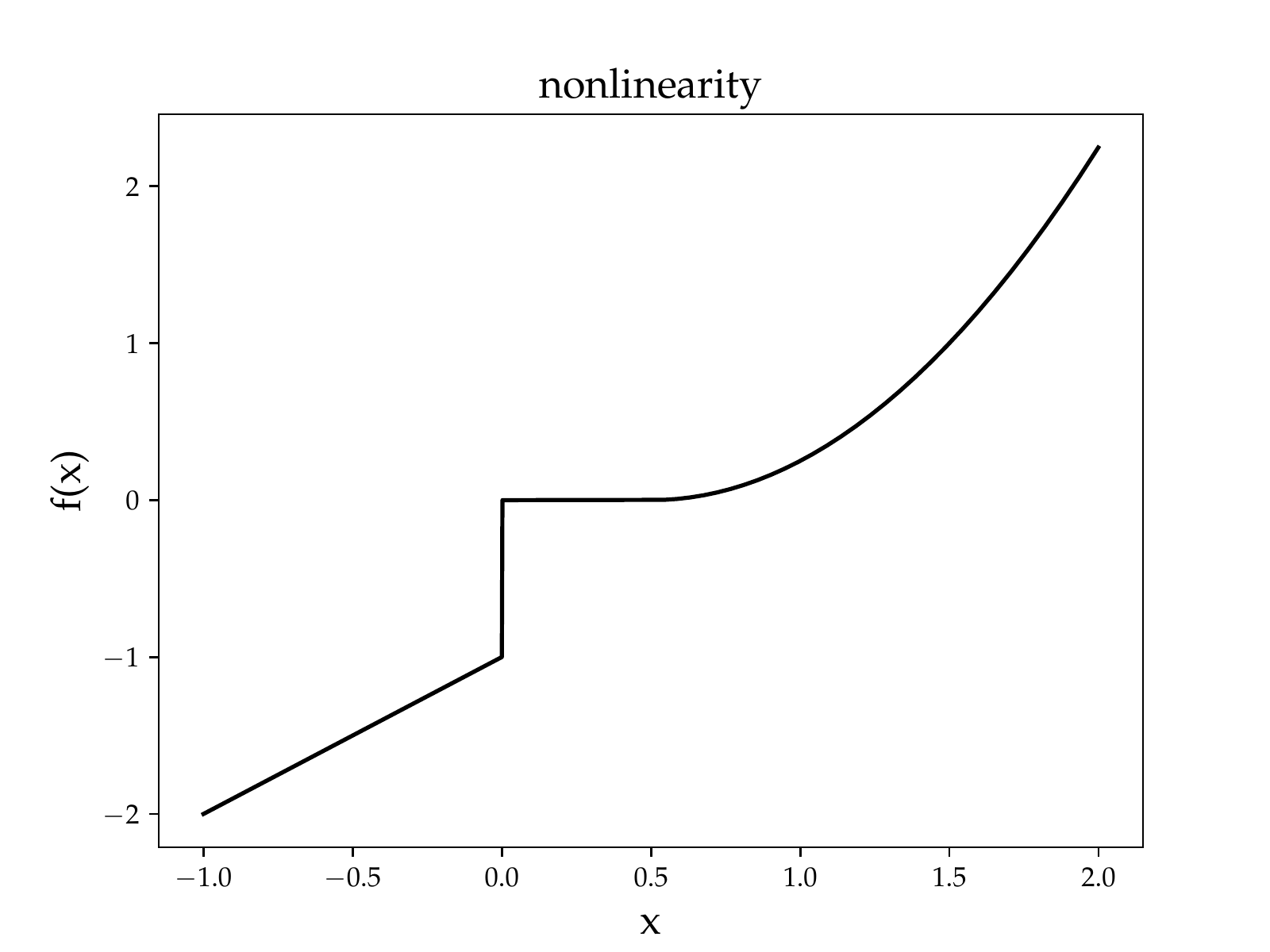}
\caption{The used nonlinearity in the first example, exhibiting linear, discontinuous, constant and quadratic parts.}
\label{fig:nonlinearity}
\end{center}
\end{figure}

\begin{figure}

\begin{center}

\includegraphics[scale = 0.5]{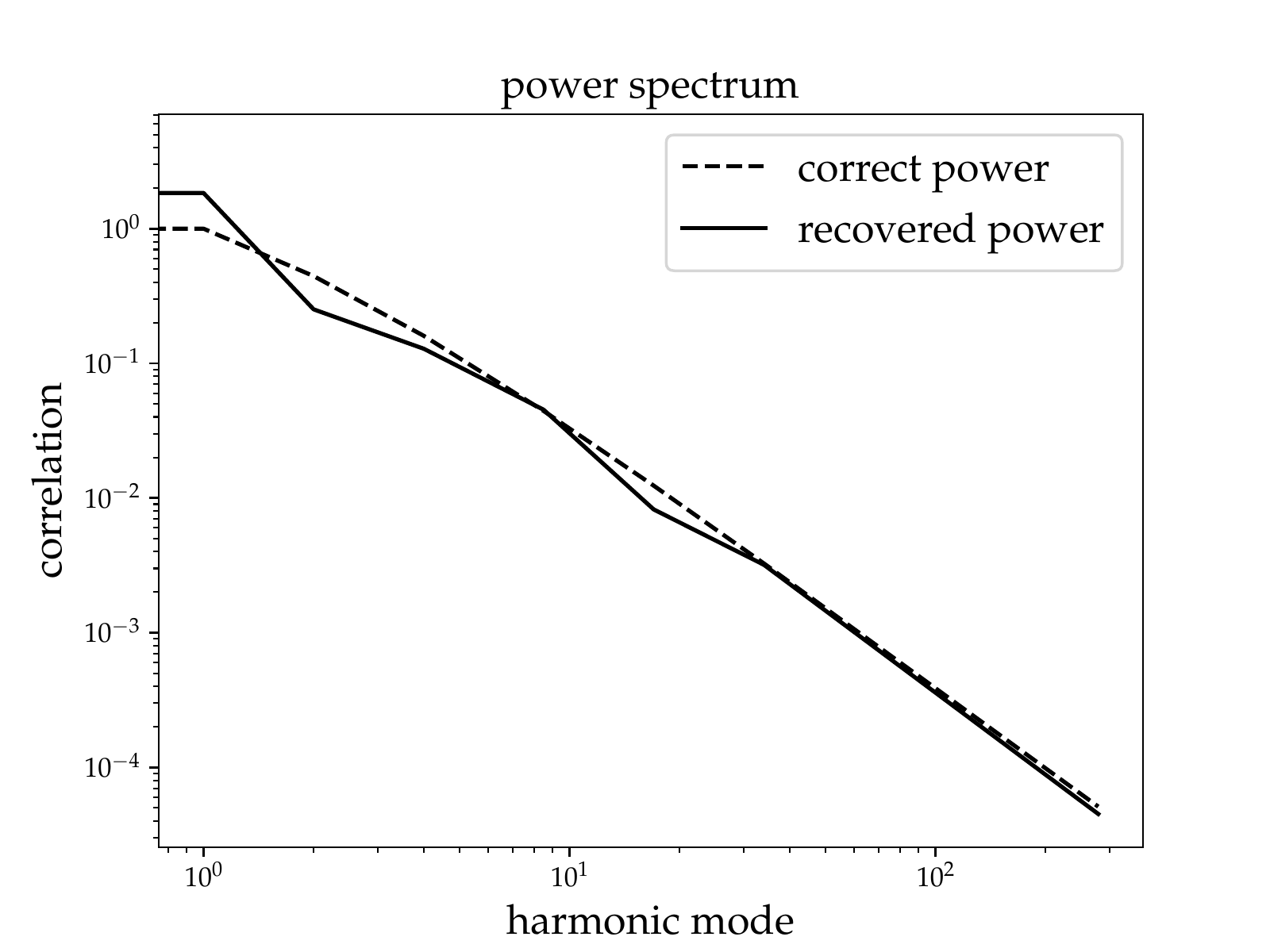}
\caption{The real underlying power spectrum of the signal together with its reconstruction.}
\label{fig:powerspectrum}

\end{center}
\end{figure}

\begin{figure}

\begin{center}

\includegraphics[scale = 0.5]{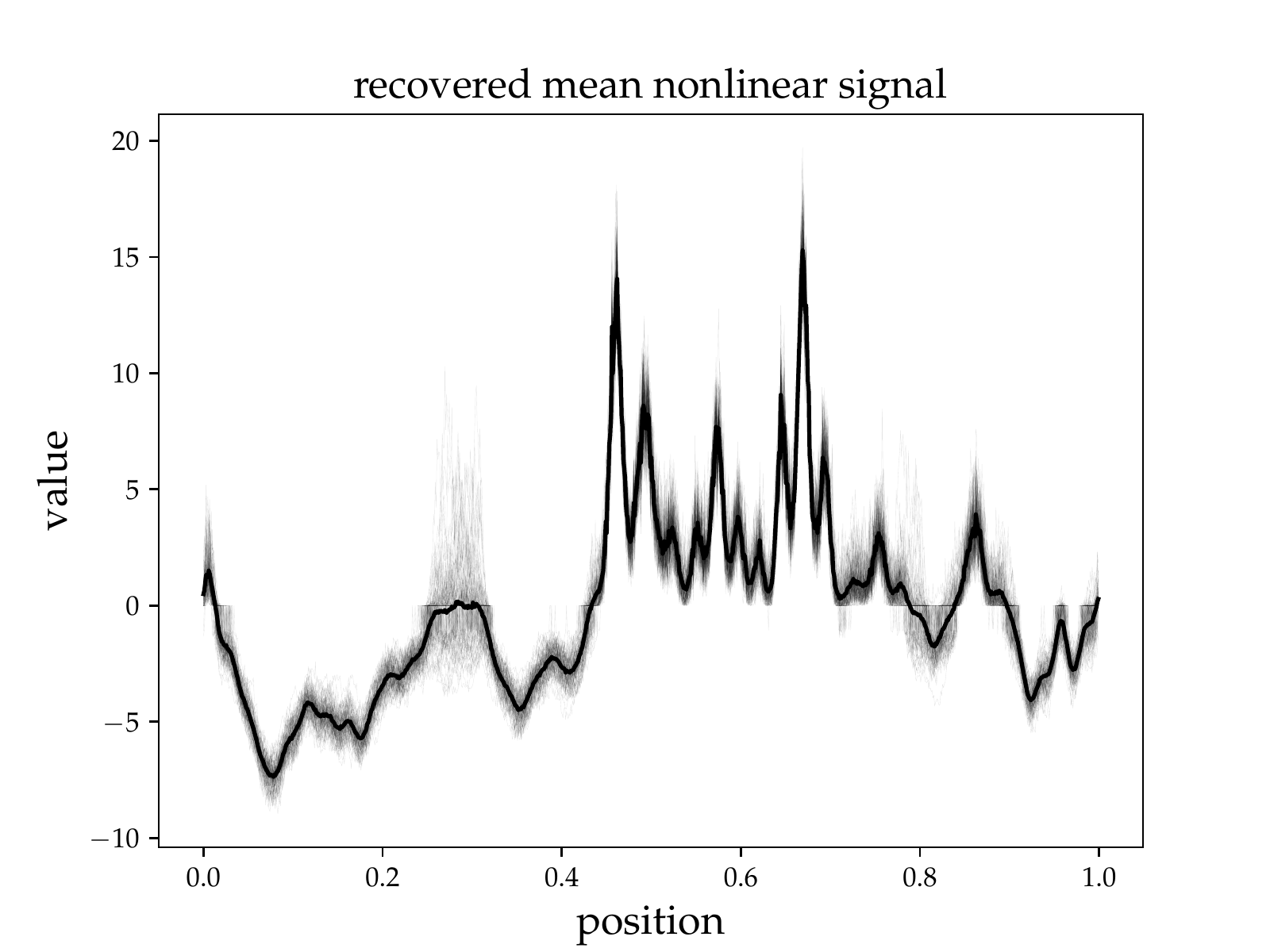}
\caption{The nonlinear response applied to the posterior samples, together with their mean. }
\label{fig:nonlinear_reconstruction}
\end{center}
\end{figure}

\begin{figure}

\begin{center}

\includegraphics[scale = 0.5]{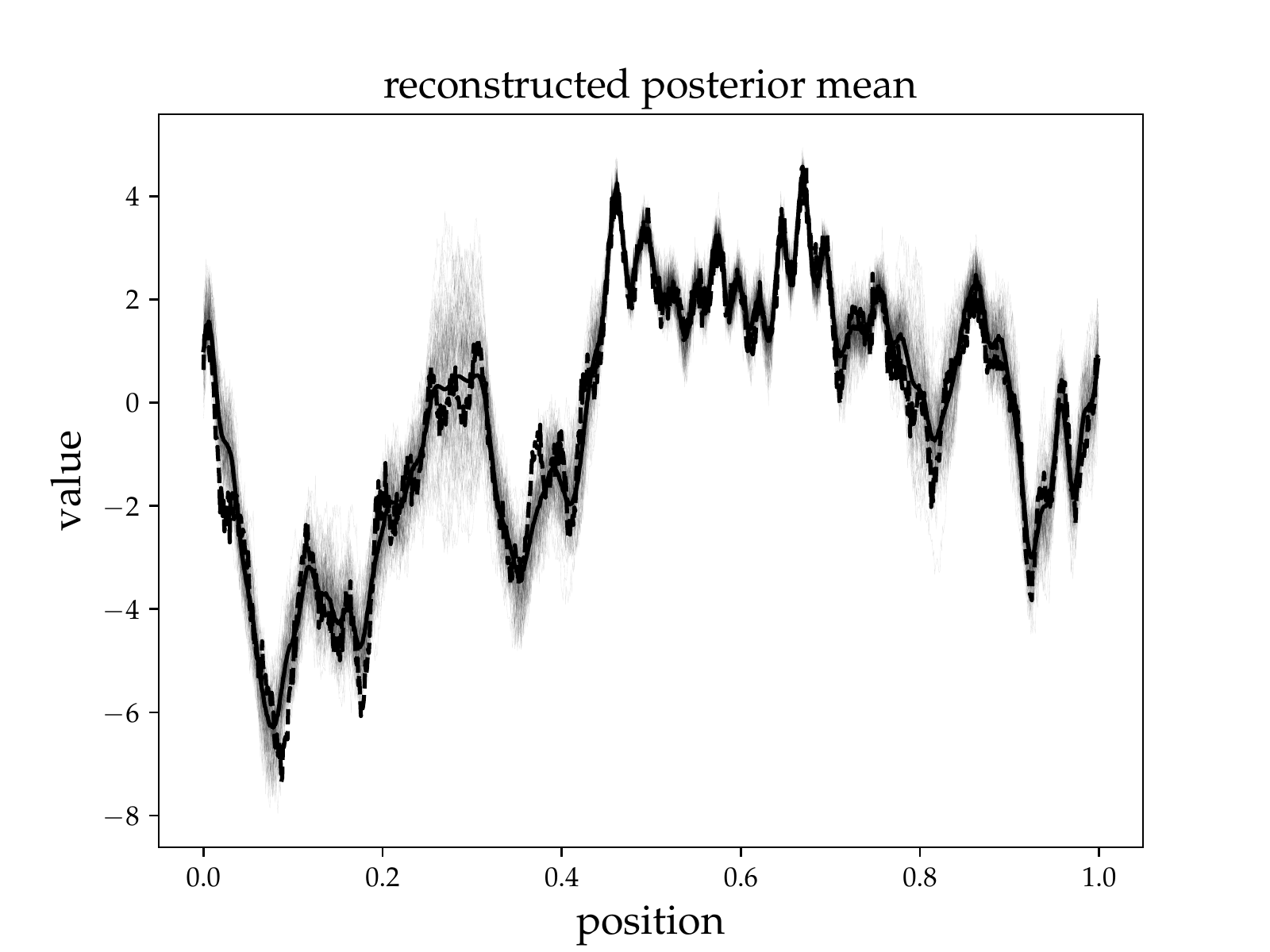}
\caption{The correct signal (dashed line), its reconstruction (solid line) and samples drawn from the posterior (grey) indicating its uncertainty variance }
\label{fig:reconstruction}

\end{center}
\end{figure}

The signal field in the first example has a resolution of $1024$ equally spaced points in one dimension. It is drawn from a Gaussian distribution with power spectrum
\begin{align}
p(k) = \frac{4}{(k+1)^2} \text{,}
\end{align}
which, as well as its reconstruction, can be seen in Fig.\ (\ref{fig:powerspectrum}). The noise covariance is $N=5 *\mathbb{1}$. The nonlinearity has the form
\begin{align}
  f(x)=\begin{cases}
    x-1 & \text{for $x<0$}\\
    0 &\text{for $0\leqslant x<\frac{1}{2}$}\\
    x^2-x + \frac{1}{4} & \text{for $\frac{1}{2} \leqslant x$,}
  \end{cases}
\end{align}
and is illustrated in Fig.\ (\ref{fig:nonlinearity}).
For negative values this function is linear. At $x=0$ this function is discontinuous, jumping from $-1$ to $0$. Between $x=0$ and $x=\frac{1}{2}$ it is constantly zero and therefore blind to any variations of the signal in this range. For larger $x$ it behaves quadratic. It does not resemble any reasonable measurement function, indeed this form is solely chosen to demonstrate the capabilities of the algorithm. Despite all those problematic properties of the response function the signal can be reconstructed. For the algorithm we do need the derivative of the function. It straightforwardly reads
\begin{align}
  f'(x)=\begin{cases}
    1 & \text{for $x<0$}\\
    \infty & \text{for $x = 0$}\\
    0 &\text{for $0 < x<\frac{1}{2}$}\\
    2x-1 & \text{for $\frac{1}{2} \leqslant x$,}
  \end{cases}
\end{align}
The diverging gradient at $x=0$ does not affect the reconstruction as the probability that the field is exactly zero vanishes.
The signal and noise were drawn from the prior distribution and the data was generated according to the data equation given in Eq.\ (\ref{eq:dataequation}). The data together with the nonlinearity applied to the signal can be seen in Fig.\ (\ref{fig:data}). For a negative signal we clearly see the linear response. In the positive region the quadratic behavior produces steep spikes and large overall values. The jump of the nonlinearity is visible around zero. There we also see plateaus, originating from the flat part of the function. We started the reconstruction with random, but almost zero initial excitation and a flat power spectrum of $p=1.8 \times 10^{-2}$. The reconstructed signal field, together with the true underlying signal and posterior samples, is shown in Fig.\ (\ref{fig:reconstruction}). The samples indicate the correlation and uncertainties of the posterior field distribution. The used nonlinearity introduces complex, highly non-Gaussian uncertainty structures, which can be seen in Fig.\ (\ref{fig:nonlinear_reconstruction}). For large, positive field values the uncertainty in the reconstruction is small due to the quadratic behavior. This reduces the effective noise as the signal is stretched out, but only linearly affected by noise in the measurement. In the negative, linear regime the uncertainty is constant, as we would expect for a Wiener filter. In the vicinity of zero, the data is not strongly conclusive, as the response is flat between $0$ and $0.5$. This leads to the increase of variance within the reconstructed signal field. The recovered nonlinear signal field interpolates smoothly between the discontinuities as the sample average is calculated. We might overestimate the error in the positive direction close to zero, as the uncertainty of the signal is expressed symmetrically in the approximation, but originates from the behavior close to zero. Overall, both the signal field and its power spectrum are recovered despite the nonlinearity. The signal variance accurately describes the reconstruction uncertainty. In Fig.\ \ref{fig:powerspectrum} we see that even on small scales the correct power spectrum is recovered. This is due to the high signal-to-noise ratio in the quadratic regime, whereby variations are magnified. The algorithm provided excellent results in this synthetic example. We will continue with a real-world application.

\subsection{Two dimensional radio-interferometry of SN 3C391}
\begin{figure}
\begin{center}
\includegraphics[scale = 0.5]{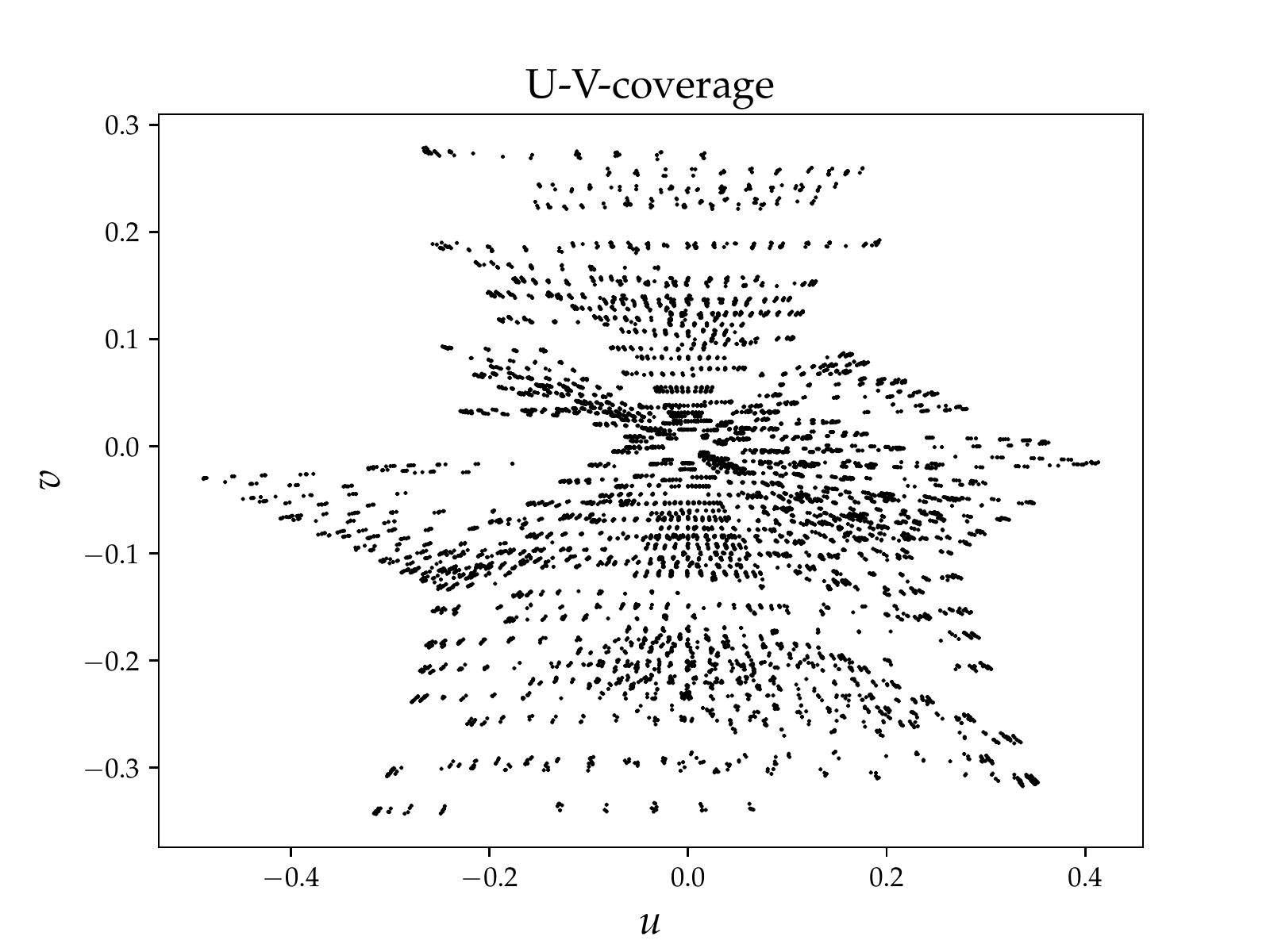}
\caption{A subset of the measured positions in the $U$-$V$ plane.}
\label{fig:uv}
\end{center}
\end{figure}

\begin{figure}
\begin{center}
\includegraphics[scale = 0.65]{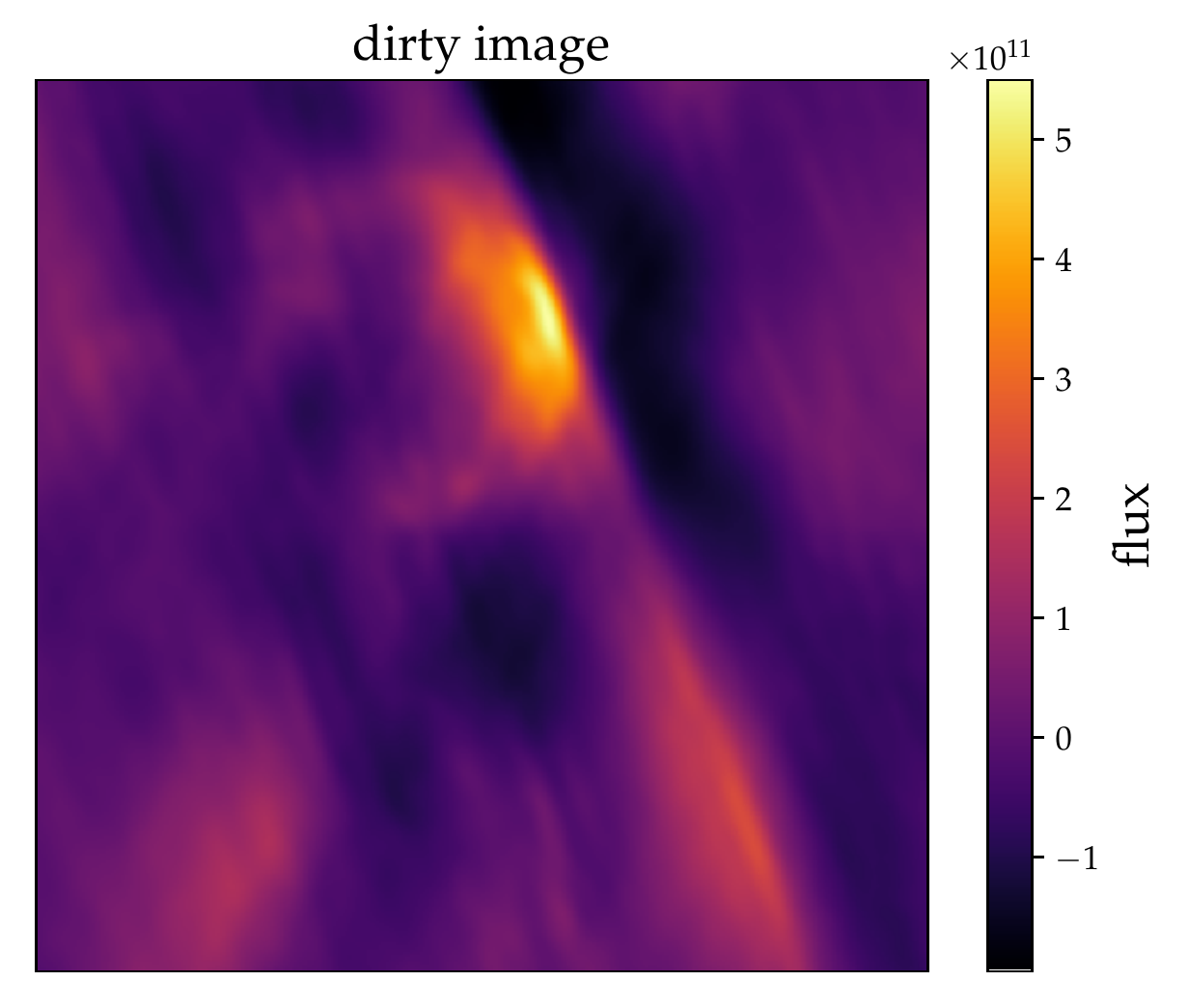}
\caption{The adjoint instrument response applied to the radio data.}
\label{fig:dirt}
\end{center}
\end{figure}

The radio emission from our universe provides us with deep insights into the nature of multiple phenomena occurring everywhere around us. Measuring and reconstructing the radio emission renders a difficult task due to the large wavelengths, terrestrial and ionospheric disturbance and incomplete coverage of the sky. To resolve extended sources radio interferometers are used. The incoming flux of multiple antennas is pairwise correlated according to a time delay corresponding to a certain position in the sky. Each antenna pair probes one position of the Fourier transformed sky brightness. The position is described by a $U$ and $V$ coordinate in the so called $U$-$V$ plane, which depend on the coordinate distance between both antennas in units of the observational wavelength. Close antenna pairs resolve large scales, while distant ones probe small image scales. The angular resolution of the image is therefore defined by the largest inter-antenna distance. The correlated antenna fluxes are called visibilities. By far not all positions in the $U$-$V$ plane, in fact typically only a small subset due to the limited amount of antennas, are probed. In our example we will look at radio interferometric data obtained by the VLA telescope of the supernova remnant SN 3C391 \citep{3C391data}, which is part of the CASA imaging software tutorial \citep{CASAtutorial}. It contains roughly $7\,000\,000$ measured points in the Fourier plane. The back projection of the data using the adjoint instrument response, also called dirty image, is shown in Fig.\ \ref{fig:dirt}. A random subset of all measured positions illustrates the $U$-$V$ coverage in Fig.\ \ref{fig:uv}. It is problematic to obtain accurate noise estimates for the data, as they depend on the instrument calibration, which can change due to environmental effects. To overcome this problem we extended our model slightly by also reconstructing the noise level from the data itself. This is done analogously to the amplitude spectrum. We parametrized the noise covariance as $N = \widehat{e^{\eta}}$ to enforce positivity and allow for exponential variation. $\eta$ is a vector of the length of the data. In addition to that we introduce a weak inverse gamma prior to slightly regulate its magnitude and to ensure numerical stability. The corresponding KL-divergence and gradient then read:

\begin{align}
\mathcal{D}_\mathrm{KL} \: \widehat{=}\: & \langle \frac{1}{2}(d-Rf( A_*\xi))^\dagger \widehat{e^{-\eta}}(d-Rf( A_*\xi))\nonumber\rangle_{\mathcal{G}(\xi-t,\Xi)} \\
& + \frac{1}{2} 1^\dagger \eta \nonumber \\
& + (\beta-1)^\dagger \eta + q^\dagger e^{-\eta}  \text {, and} \\
\frac{\delta \mathcal{D}_\mathrm{KL}}{\delta \eta} \: \widehat{=}\: & - \langle \frac{1}{2}(d-Rf( A_*\xi))^\dagger \widehat{\widehat{e^{-\eta}}}(d-Rf( A_*\xi))\nonumber\rangle_{\mathcal{G}(\xi-t,\Xi)} \\
& + \frac{1}{2} +(1-\beta) - q \widehat{e^{-\eta}}  \text{.}
\end{align}
To infer the noise covariance, we cannot drop the noise normalization term $\frac{1}{2}\mathrm{ln}\vert N\vert = \frac{1}{2} 1^\dagger \eta$ from the Hamiltonian, as it is no longer a constant. The prior parameters $\beta$ and $q$ describe the shape and scale of the inverse gamma prior on $\eta$, respectively. The inference procedure adapts to this by simply adding a minimization step with respect to the noise covariance in each iteration. The noise is therefore continuously calibrated to the current reconstruction.

The chosen nonlinearity in this example is the exponential function, which leads to a log-normal model for the flux. It allows for exponential brightness variations across the sky and enforces positivity of the underlying intensity distribution. This model has already been successfully applied to radio interferometric data sets in the classical critical filter formulation, but always suffered from slow convergence \citep{RESOLVE}.

We initialized this reconstruction with a flat amplitude spectrum and a close to zero, white excitation and the data variance as initial noise estimate. The strength of the smoothness prior for the amplitudes was set to $\sigma = 1$, allowing for deviations from smooth spectra of one per magnitude. $\beta$ was set to $2.0000002$ and $q$ to $2 * 10^{-5}$. This leads to a mode of the inverse gamma distribution at $10^{-5}$, which corresponds roughly to the data variance, and a mean at $ 100$. The mean had to be that high to not restrict the noise too much. Due to the heavy tails of the inverse gamma distribution the relation between mode and mean is rather counter-intuitive. The choice of these values only slightly effects the reconstruction. The main purpose is to avoid numerical instability by divergence of individual points.

\begin{figure}
\begin{center}

\includegraphics[scale = 0.65]{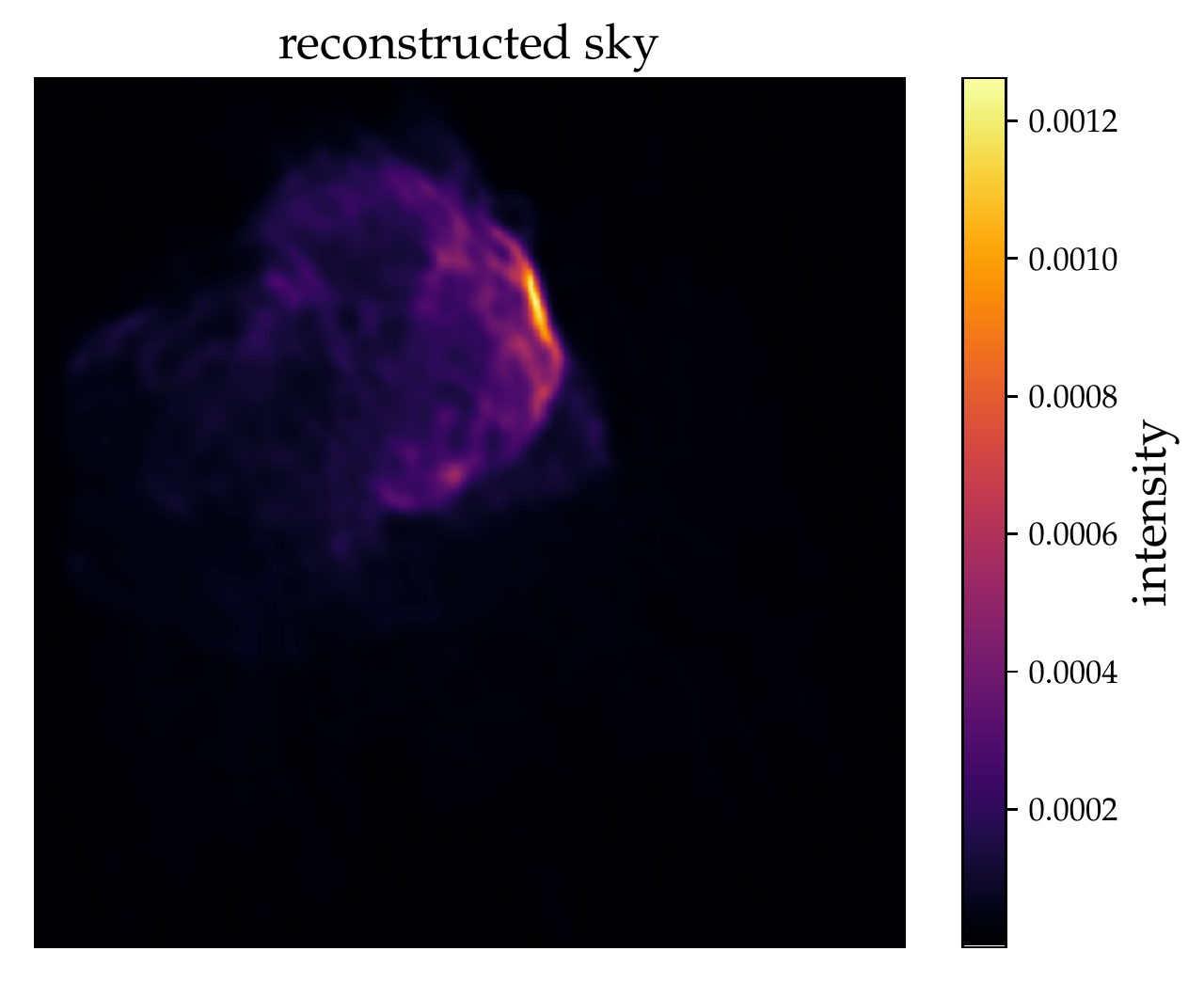}
\caption{The reconstruction of the sky radio intensity of SN 3C391. }
\label{fig:sky}
\end{center}
\end{figure}

\begin{figure}
\begin{center}
\includegraphics[scale = 0.5]{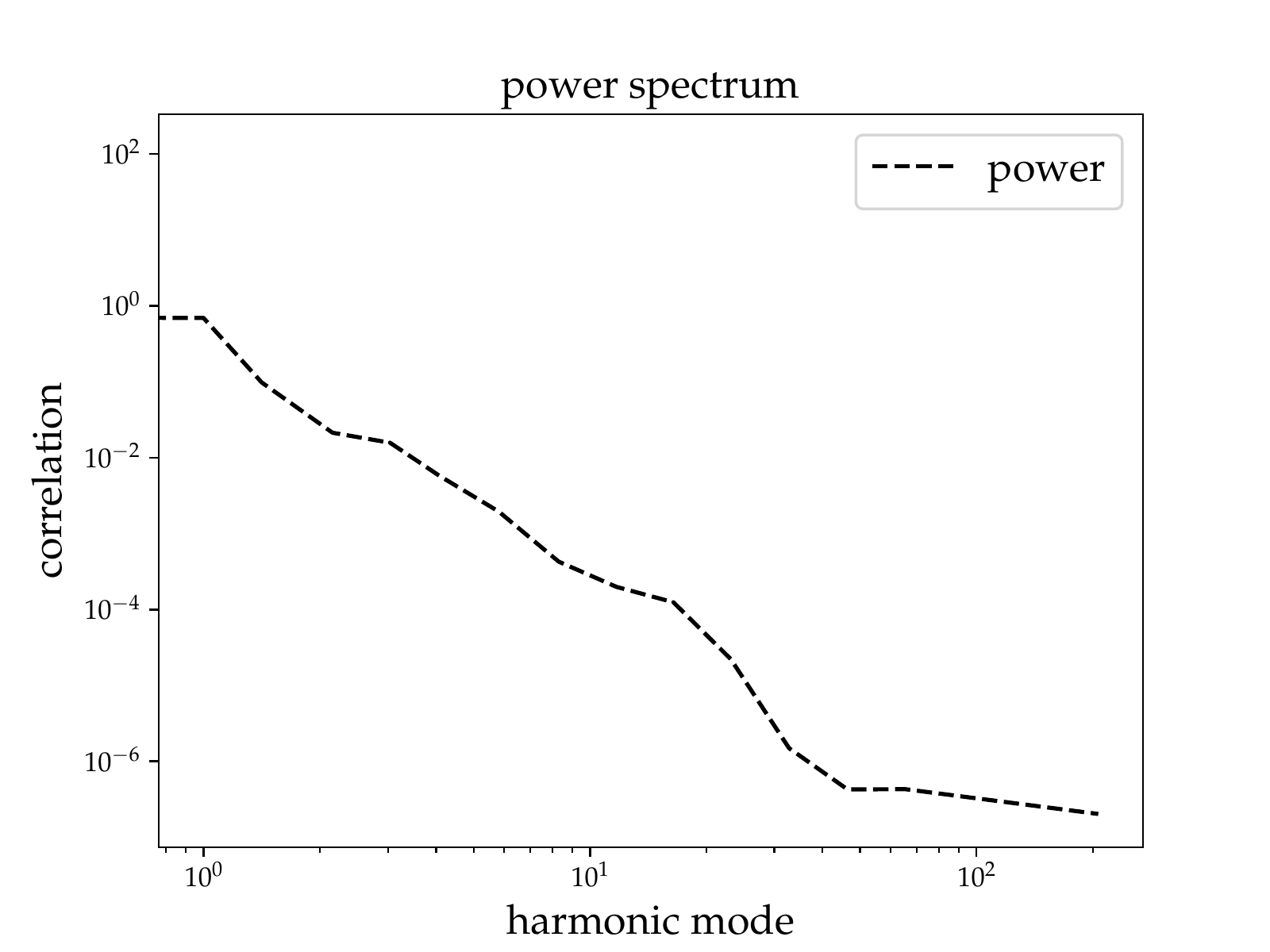}
\caption{The reconstructed correlation structure of SN 3C391. }
\label{fig:3C391_power}
\end{center}
\end{figure}

\begin{figure}
\begin{center}
\includegraphics[scale = 0.65]{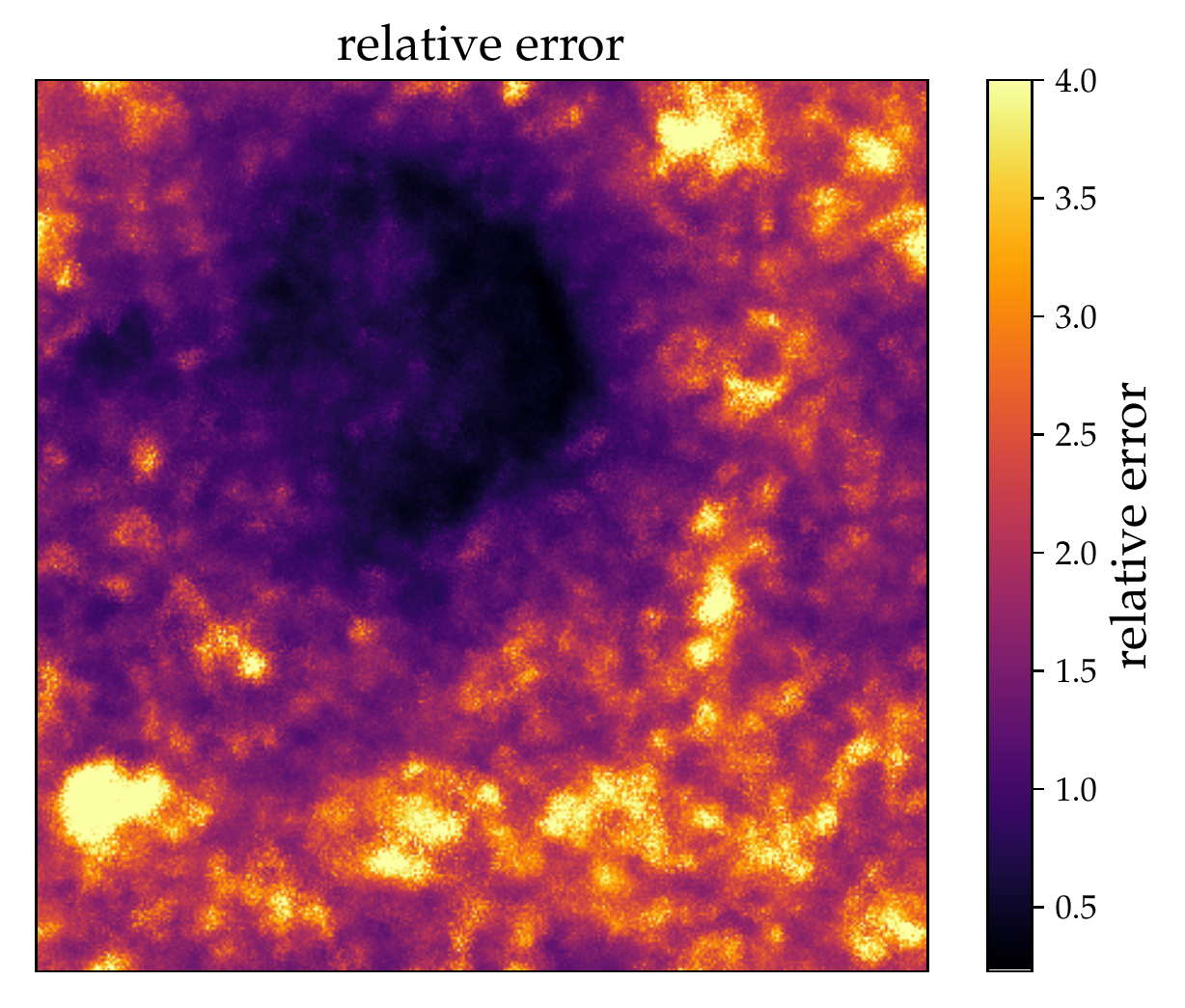}
\caption{The sampled relative error of Fig (\ref{fig:sky}), quantifying the fidelity of the reconstruction in certain regions.}
\label{fig:error}
\end{center}
\end{figure}

The result of the reconstruction and the recovered correlation structure can be seen in Fig.\ \ref{fig:sky} and Fig.\ \ref{fig:3C391_power}, respectively. We sharply resolve the shock front of the supernova and surrounding filamentary structures, even for low intensities. From posterior samples we can also give an estimate of the uncertainty of the resolved structures. The relative uncertainty defined as
\begin{align}
E = \frac{\sqrt{\langle\left( e^s - \langle e^s \rangle\right)^2 \rangle }}{\langle e^s \rangle}
\end{align} 
is shown in Fig \ref{fig:error}, using $100$ posterior samples. We clearly see high relative uncertainty in regions with low intensity and vice versa. Using only a limited number of samples leads to some remaining sampling artifacts, but for larger scales the error estimate is reliable. Overall the reconstruction of the supernova remnant SN 3C391 using the presented algorithm gave satisfying results, allowing for highly resolved radio reconstructions, even without having the instrument's noise levels available.

\section{Conclusion}
We presented a method to reconstruct autocorrelated signals from nonlinear, noisy measurements without knowing the correlation a priori, but rather reconstructing it simultaneously. Reformulating the mathematical foundations of the underlying model results in a drastic speed up compared the previous approach, enabling whole new fields of application. By using approximate posterior samples one can consider complex uncertainty structures, allowing to generalize the method to arbitrary monotonous nonlinearities. 

These can be used to model nonlinear measurement instruments or impose constraints on the signal of interest. Uncertainty information about arbitrary posterior quantities can be obtained by using the approximate posterior samples. We demonstrated the applicability of the algorithm in the context of highly different imaging problems with diverse instruments, nonlinear characteristics and dimension. This can be achieved by formulating the model using the abstract language of information field theory.

 As shown in the second example of reconstructing the supernova remnant, the inference of additional unknown quantities such as the noise covariance can be included easily. Adaption to further instruments is straightforward, given a decent description of the instrument response $R$. The computational effort is still large compared to simpler methods, but the obtained results greatly surpass them in quality, especially in the high noise regime.

\section{Acknowledgment}
We acknowledge Philipp Arras, Philipp Frank, Reimar Leike and Martin Reinecke for fruitful discussions and remarks on the manuscript, as well as the Computational Center for Particle and Astrophysics (C2PAP) for providing computational resources and support.

\bibliographystyle{apsrev4-1}

\bibliography{citations}

\begin{thebibliography}{20}%
\makeatletter
\providecommand \@ifxundefined [1]{%
 \@ifx{#1\undefined}
}%
\providecommand \@ifnum [1]{%
 \ifnum #1\expandafter \@firstoftwo
 \else \expandafter \@secondoftwo
 \fi
}%
\providecommand \@ifx [1]{%
 \ifx #1\expandafter \@firstoftwo
 \else \expandafter \@secondoftwo
 \fi
}%
\providecommand \natexlab [1]{#1}%
\providecommand \enquote  [1]{``#1''}%
\providecommand \bibnamefont  [1]{#1}%
\providecommand \bibfnamefont [1]{#1}%
\providecommand \citenamefont [1]{#1}%
\providecommand \href@noop [0]{\@secondoftwo}%
\providecommand \href [0]{\begingroup \@sanitize@url \@href}%
\providecommand \@href[1]{\@@startlink{#1}\@@href}%
\providecommand \@@href[1]{\endgroup#1\@@endlink}%
\providecommand \@sanitize@url [0]{\catcode `\\12\catcode `\$12\catcode
  `\&12\catcode `\#12\catcode `\^12\catcode `\_12\catcode `\%12\relax}%
\providecommand \@@startlink[1]{}%
\providecommand \@@endlink[0]{}%
\providecommand \url  [0]{\begingroup\@sanitize@url \@url }%
\providecommand \@url [1]{\endgroup\@href {#1}{\urlprefix }}%
\providecommand \urlprefix  [0]{URL }%
\providecommand \Eprint [0]{\href }%
\providecommand \doibase [0]{http://dx.doi.org/}%
\providecommand \selectlanguage [0]{\@gobble}%
\providecommand \bibinfo  [0]{\@secondoftwo}%
\providecommand \bibfield  [0]{\@secondoftwo}%
\providecommand \translation [1]{[#1]}%
\providecommand \BibitemOpen [0]{}%
\providecommand \bibitemStop [0]{}%
\providecommand \bibitemNoStop [0]{.\EOS\space}%
\providecommand \EOS [0]{\spacefactor3000\relax}%
\providecommand \BibitemShut  [1]{\csname bibitem#1\endcsname}%
\let\auto@bib@innerbib\@empty
\bibitem [{\citenamefont {Deans}(2007)}]{filteredbackprojection}%
  \BibitemOpen
  \bibfield  {author} {\bibinfo {author} {\bibfnamefont {S.~R.}\ \bibnamefont
  {Deans}},\ }\href@noop {} {\emph {\bibinfo {title} {The Radon transform and
  some of its applications}}}\ (\bibinfo  {publisher} {Courier Corporation},\
  \bibinfo {year} {2007})\BibitemShut {NoStop}%
\bibitem [{\citenamefont {Donoho}(2006)}]{compressed}%
  \BibitemOpen
  \bibfield  {author} {\bibinfo {author} {\bibfnamefont {D.~L.}\ \bibnamefont
  {Donoho}},\ }\href@noop {} {\bibfield  {journal} {\bibinfo  {journal} {IEEE
  Transactions on information theory}\ }\textbf {\bibinfo {volume} {52}},\
  \bibinfo {pages} {1289} (\bibinfo {year} {2006})}\BibitemShut {NoStop}%
\bibitem [{\citenamefont {Cand{\`e}s}\ \emph {et~al.}(2006)\citenamefont
  {Cand{\`e}s}, \citenamefont {Romberg},\ and\ \citenamefont {Tao}}]{robustCS}%
  \BibitemOpen
  \bibfield  {author} {\bibinfo {author} {\bibfnamefont {E.~J.}\ \bibnamefont
  {Cand{\`e}s}}, \bibinfo {author} {\bibfnamefont {J.}~\bibnamefont {Romberg}},
  \ and\ \bibinfo {author} {\bibfnamefont {T.}~\bibnamefont {Tao}},\
  }\href@noop {} {\bibfield  {journal} {\bibinfo  {journal} {IEEE Transactions
  on information theory}\ }\textbf {\bibinfo {volume} {52}},\ \bibinfo {pages}
  {489} (\bibinfo {year} {2006})}\BibitemShut {NoStop}%
\bibitem [{\citenamefont {Nyquist}(1928)}]{nyquisttelegraph}%
  \BibitemOpen
  \bibfield  {author} {\bibinfo {author} {\bibfnamefont {H.}~\bibnamefont
  {Nyquist}},\ }\href@noop {} {\bibfield  {journal} {\bibinfo  {journal}
  {Transactions of the American Institute of Electrical Engineers}\ }\textbf
  {\bibinfo {volume} {47}},\ \bibinfo {pages} {617} (\bibinfo {year}
  {1928})}\BibitemShut {NoStop}%
\bibitem [{\citenamefont {Shannon}(1949)}]{shannoncommunication}%
  \BibitemOpen
  \bibfield  {author} {\bibinfo {author} {\bibfnamefont {C.~E.}\ \bibnamefont
  {Shannon}},\ }\href@noop {} {\bibfield  {journal} {\bibinfo  {journal}
  {Proceedings of the IRE}\ }\textbf {\bibinfo {volume} {37}},\ \bibinfo
  {pages} {10} (\bibinfo {year} {1949})}\BibitemShut {NoStop}%
\bibitem [{\citenamefont {Wiener}(1949)}]{Wiener}%
  \BibitemOpen
  \bibfield  {author} {\bibinfo {author} {\bibfnamefont {N.}~\bibnamefont
  {Wiener}},\ }\href@noop {} {\emph {\bibinfo {title} {Extrapolation,
  interpolation, and smoothing of stationary time series}}},\ Vol.~\bibinfo
  {volume} {2}\ (\bibinfo  {publisher} {MIT press Cambridge},\ \bibinfo {year}
  {1949})\BibitemShut {NoStop}%
\bibitem [{\citenamefont {En{\ss}lin}\ \emph {et~al.}(2009)\citenamefont
  {En{\ss}lin}, \citenamefont {Frommert},\ and\ \citenamefont {Kitaura}}]{IFT}%
  \BibitemOpen
  \bibfield  {author} {\bibinfo {author} {\bibfnamefont {T.~A.}\ \bibnamefont
  {En{\ss}lin}}, \bibinfo {author} {\bibfnamefont {M.}~\bibnamefont
  {Frommert}}, \ and\ \bibinfo {author} {\bibfnamefont {F.~S.}\ \bibnamefont
  {Kitaura}},\ }\href@noop {} {\bibfield  {journal} {\bibinfo  {journal}
  {Physical Review D}\ }\textbf {\bibinfo {volume} {80}},\ \bibinfo {pages}
  {105005} (\bibinfo {year} {2009})}\BibitemShut {NoStop}%
\bibitem [{\citenamefont {Wandelt}(2004)}]{MAGIC}%
  \BibitemOpen
  \bibfield  {author} {\bibinfo {author} {\bibfnamefont {B.~D.}\ \bibnamefont
  {Wandelt}},\ }\href@noop {} {\bibfield  {journal} {\bibinfo  {journal} {arXiv
  preprint astro-ph/0401623}\ } (\bibinfo {year} {2004})}\BibitemShut {NoStop}%
\bibitem [{\citenamefont {En{\ss}lin}\ and\ \citenamefont
  {Frommert}(2011)}]{ensslinfrommert}%
  \BibitemOpen
  \bibfield  {author} {\bibinfo {author} {\bibfnamefont {T.~A.}\ \bibnamefont
  {En{\ss}lin}}\ and\ \bibinfo {author} {\bibfnamefont {M.}~\bibnamefont
  {Frommert}},\ }\href@noop {} {\bibfield  {journal} {\bibinfo  {journal}
  {Physical Review D}\ }\textbf {\bibinfo {volume} {83}},\ \bibinfo {pages}
  {105014} (\bibinfo {year} {2011})}\BibitemShut {NoStop}%
\bibitem [{\citenamefont {Selig}\ \emph {et~al.}(2015)\citenamefont {Selig},
  \citenamefont {Vacca}, \citenamefont {Oppermann},\ and\ \citenamefont
  {En{\ss}lin}}]{D3PO}%
  \BibitemOpen
  \bibfield  {author} {\bibinfo {author} {\bibfnamefont {M.}~\bibnamefont
  {Selig}}, \bibinfo {author} {\bibfnamefont {V.}~\bibnamefont {Vacca}},
  \bibinfo {author} {\bibfnamefont {N.}~\bibnamefont {Oppermann}}, \ and\
  \bibinfo {author} {\bibfnamefont {T.~A.}\ \bibnamefont {En{\ss}lin}},\
  }\href@noop {} {\bibfield  {journal} {\bibinfo  {journal} {Astronomy \&
  Astrophysics}\ }\textbf {\bibinfo {volume} {581}},\ \bibinfo {pages} {A126}
  (\bibinfo {year} {2015})}\BibitemShut {NoStop}%
\bibitem [{\citenamefont {Junklewitz}\ \emph {et~al.}(2016)\citenamefont
  {Junklewitz}, \citenamefont {Bell}, \citenamefont {Selig},\ and\
  \citenamefont {En{\ss}lin}}]{RESOLVE}%
  \BibitemOpen
  \bibfield  {author} {\bibinfo {author} {\bibfnamefont {H.}~\bibnamefont
  {Junklewitz}}, \bibinfo {author} {\bibfnamefont {M.}~\bibnamefont {Bell}},
  \bibinfo {author} {\bibfnamefont {M.}~\bibnamefont {Selig}}, \ and\ \bibinfo
  {author} {\bibfnamefont {T.}~\bibnamefont {En{\ss}lin}},\ }\href@noop {}
  {\bibfield  {journal} {\bibinfo  {journal} {Astronomy \& Astrophysics}\
  }\textbf {\bibinfo {volume} {586}},\ \bibinfo {pages} {A76} (\bibinfo {year}
  {2016})}\BibitemShut {NoStop}%
\bibitem [{\citenamefont {Kullback}\ and\ \citenamefont
  {Leibler}(1951)}]{KLdivergence}%
  \BibitemOpen
  \bibfield  {author} {\bibinfo {author} {\bibfnamefont {S.}~\bibnamefont
  {Kullback}}\ and\ \bibinfo {author} {\bibfnamefont {R.~A.}\ \bibnamefont
  {Leibler}},\ }\href@noop {} {\bibfield  {journal} {\bibinfo  {journal} {The
  annals of mathematical statistics}\ }\textbf {\bibinfo {volume} {22}},\
  \bibinfo {pages} {79} (\bibinfo {year} {1951})}\BibitemShut {NoStop}%
\bibitem [{\citenamefont {Oppermann}\ \emph {et~al.}(2013)\citenamefont
  {Oppermann}, \citenamefont {Selig}, \citenamefont {Bell},\ and\ \citenamefont
  {En{\ss}lin}}]{smoothpower}%
  \BibitemOpen
  \bibfield  {author} {\bibinfo {author} {\bibfnamefont {N.}~\bibnamefont
  {Oppermann}}, \bibinfo {author} {\bibfnamefont {M.}~\bibnamefont {Selig}},
  \bibinfo {author} {\bibfnamefont {M.~R.}\ \bibnamefont {Bell}}, \ and\
  \bibinfo {author} {\bibfnamefont {T.~A.}\ \bibnamefont {En{\ss}lin}},\
  }\href@noop {} {\bibfield  {journal} {\bibinfo  {journal} {Physical Review
  E}\ }\textbf {\bibinfo {volume} {87}},\ \bibinfo {pages} {032136} (\bibinfo
  {year} {2013})}\BibitemShut {NoStop}%
\bibitem [{\citenamefont {Steininger}\ \emph {et~al.}(2017)\citenamefont
  {Steininger}, \citenamefont {Dixit}, \citenamefont {Frank}, \citenamefont
  {Greiner}, \citenamefont {Hutschenreuter}, \citenamefont {Knollm{\"u}ller},
  \citenamefont {Leike}, \citenamefont {Porqueres}, \citenamefont {Pumpe},
  \citenamefont {Reinecke} \emph {et~al.}}]{Nifty3}%
  \BibitemOpen
  \bibfield  {author} {\bibinfo {author} {\bibfnamefont {T.}~\bibnamefont
  {Steininger}}, \bibinfo {author} {\bibfnamefont {J.}~\bibnamefont {Dixit}},
  \bibinfo {author} {\bibfnamefont {P.}~\bibnamefont {Frank}}, \bibinfo
  {author} {\bibfnamefont {M.}~\bibnamefont {Greiner}}, \bibinfo {author}
  {\bibfnamefont {S.}~\bibnamefont {Hutschenreuter}}, \bibinfo {author}
  {\bibfnamefont {J.}~\bibnamefont {Knollm{\"u}ller}}, \bibinfo {author}
  {\bibfnamefont {R.}~\bibnamefont {Leike}}, \bibinfo {author} {\bibfnamefont
  {N.}~\bibnamefont {Porqueres}}, \bibinfo {author} {\bibfnamefont
  {D.}~\bibnamefont {Pumpe}}, \bibinfo {author} {\bibfnamefont
  {M.}~\bibnamefont {Reinecke}},  \emph {et~al.},\ }\href@noop {} {\bibfield
  {journal} {\bibinfo  {journal} {arXiv preprint arXiv:1708.01073}\ } (\bibinfo
  {year} {2017})}\BibitemShut {NoStop}%
\bibitem [{\citenamefont {Miller-Jones}(2010)}]{3C391data}%
  \BibitemOpen
  \bibfield  {author} {\bibinfo {author} {\bibfnamefont {J.}~\bibnamefont
  {Miller-Jones}},\ }\href@noop {} {\enquote {\bibinfo {title}
  {\path{2010-04-24_0801_TDEM0001}},}\ }\bibinfo {howpublished} {NRAO Science
  Data Archive} (\bibinfo {year} {2010})\BibitemShut {NoStop}%
\bibitem [{CAS()}]{CASAtutorial}%
  \BibitemOpen
  \href@noop {} {\enquote {\bibinfo {title} {\path{VLA Continuum Tutorial
  3C391-CASA5.0.0}},}\ }\bibinfo {howpublished}
  {\url{https://casaguides.nrao.edu/index.php/VLA_Continuum_Tutorial_3C391-CASA5.0.0}},\
  \bibinfo {note} {accessed: 26.10.2017}\BibitemShut {NoStop}%
\bibitem [{\citenamefont {Jaynes}(1957)}]{jaynes}%
  \BibitemOpen
  \bibfield  {author} {\bibinfo {author} {\bibfnamefont {E.~T.}\ \bibnamefont
  {Jaynes}},\ }\href@noop {} {\bibfield  {journal} {\bibinfo  {journal}
  {Physical review}\ }\textbf {\bibinfo {volume} {106}},\ \bibinfo {pages}
  {620} (\bibinfo {year} {1957})}\BibitemShut {NoStop}%
\bibitem [{\citenamefont {Khintchin}(1934)}]{Khintchin}%
  \BibitemOpen
  \bibfield  {author} {\bibinfo {author} {\bibfnamefont {A.}~\bibnamefont
  {Khintchin}},\ }\href@noop {} {\bibfield  {journal} {\bibinfo  {journal}
  {Mathematische Annalen}\ }\textbf {\bibinfo {volume} {109}},\ \bibinfo
  {pages} {604} (\bibinfo {year} {1934})}\BibitemShut {NoStop}%
\bibitem [{\citenamefont {En{\ss}lin}\ and\ \citenamefont
  {Weig}(2010)}]{ensslinweig}%
  \BibitemOpen
  \bibfield  {author} {\bibinfo {author} {\bibfnamefont {T.~A.}\ \bibnamefont
  {En{\ss}lin}}\ and\ \bibinfo {author} {\bibfnamefont {C.}~\bibnamefont
  {Weig}},\ }\href@noop {} {\bibfield  {journal} {\bibinfo  {journal} {Physical
  Review E}\ }\textbf {\bibinfo {volume} {82}},\ \bibinfo {pages} {051112}
  (\bibinfo {year} {2010})}\BibitemShut {NoStop}%
\bibitem [{\citenamefont {Knollm{\"u}ller}\ and\ \citenamefont
  {En{\ss}lin}(2017)}]{NICAAC}%
  \BibitemOpen
  \bibfield  {author} {\bibinfo {author} {\bibfnamefont {J.}~\bibnamefont
  {Knollm{\"u}ller}}\ and\ \bibinfo {author} {\bibfnamefont {T.~A.}\
  \bibnamefont {En{\ss}lin}},\ }\href@noop {} {\bibfield  {journal} {\bibinfo
  {journal} {Physical Review E}\ }\textbf {\bibinfo {volume} {96}},\ \bibinfo
  {pages} {042114} (\bibinfo {year} {2017})}\BibitemShut {NoStop}%
\end{thebibliography}%

\appendix

\section{Wiener Filter}
\label{sec:WienerFilter}
We start with a description of the measurement equation, which establishes a relation between the physical signal and the data we obtained from the measurement process. The signal should be a physical field with infinite resolution, at least in theory. This allows us to apply the rich toolbox of IFT \citep{IFT}, which applies information theory to fields. The data itself always has to be discrete, as we can never store the infinite degrees of freedom of the field, which corresponds to infinite information. We therefore have to use some instrument which probes the field in some way, for example by integrating it in some area and thereby performing a discretization. Additionally, more complex instrument characteristics can be modeled, as we will demonstrate in the examples later. Finally, in this work we consider additive, Gaussian noise on the data.

Mathematically we can express the result of any linear measurement as data $d$ being the outcome of applying the measurement instrument's response $R$ to the physical field $s$, transforming it into a discrete quantity, which is finally corrupted by instrument noise $n$:
\begin{align}
d = R s + n
\end{align}
As mentioned, we will assume Gaussian noise $n$ with a known covariance structure $N$:
\begin{align}
n  \curvearrowleft \mathcal{G}(n,N) \text {, with}\\
\label{eq:gaussian}
\mathcal{G}(n,N) \equiv \frac{1}{\vert2\pi N\vert^{\frac{1}{2}}} e^{-\frac{1}{2} n^\dagger N^{-1}n}
\end{align}
Here $n^\dagger$ is the transposed and complex conjugated noise vector $n$. The covariance $N$ has the form of a matrix and the expression is therefore a multivariate Gaussian distribution.

The model likelihood describes how likely the data is, given certain quantities. In our case, the likelihood for the data given the signal and noise realization is described by a delta distribution, as all degrees of freedom are constrained, therefore
\begin{align}
\mathcal{P}(d\vert s,n) = \delta\left(d-(Rs+n)\right) \text{.} 
\end{align}
However, we are not interested in the realization of the noise and we will remove it from the formalism by marginalizing it out using our Gaussian noise model. The marginalization is performed over all possible noise configurations, denoted by the integral with respect to  $ \mathcal{D}n$.
\begin{align}
\mathcal{P}(d\vert s) =& \int \mathcal{D}n\: \mathcal{P}(d,n\vert s) = \int \mathcal{D}\: \mathcal{P}(d\vert s,n) \mathcal{P}(n) \\
=& \int  \mathcal{D} \: \delta[d-(Rs+n)] \mathcal{G}(n,N)\\
=& \: \mathcal{G}(d-Rs,N)
\end{align}
We actually want to know the probability of a signal given the data, viz.\ \emph{What does the data tell us about the signal?}. To invert the above likelihood we apply Bayes' theorem:
\begin{align}
\mathcal{P}(s\vert d) = \frac{\mathcal{P}(d\vert s) \mathcal{P}(s)}{\mathcal{P}(d)}
\end{align}
To calculate this expression we have to specify our prior distribution $\mathcal{P}(s)$ over the signal field $s$. This is a probability distribution over the space of all possible fields, which has an infinite number of degrees of freedom. For now, we assume the signal to vary around zero and to exhibit some correlation structure $S(x,x')$, which is encoded in the correlation operator $S$. With only this constraint the least informative \citep{jaynes} prior distribution is a zero centered Gaussian prior of the form
\begin{align}
\mathcal{P}(s) = \mathcal{G}(s,S) \text{.}
\end{align}
The structure of this expression is equivalent to Eq.\ \ref{eq:gaussian} if we identify the scalar product of two fields $a$ and $b$
\begin{align}
a^\dagger b = \int \mathcal{D}x \: a^*(x) b(x)
\end{align}
and apply operators via
\begin{align}
Ab = \int  \mathcal{D}x \: A(x,x') \: b(x) \text{.}
\end{align}
In analogy to statistical physics we can now reformulate Bayes' theorem by introducing the information Hamiltonian $\mathcal{H}(s,d)$ and partition function $\mathcal{Z}(d)$
\begin{align}
\mathcal{P}(s\vert d) =& \frac{1}{\mathcal{Z}(d)}e^{-\mathcal{H}(s, d)} \text {, with} \\
\mathcal{H}(s,d) \equiv& -\mathrm{ln}\left(\mathcal{P}(s,d)\right) \text {, and} \\
\mathcal{Z}(d) \equiv&  \int  \mathcal{D}s \: e^{-\mathcal{H}(s,d)} = \mathcal{P}(d) \text{.}
\end{align}
Compared to the probability distributions, the Hamiltonian behaves additive, instead of multiplicative, which makes calculations easier. For our current situation we can take the logarithms of the Gaussian likelihood and prior and add them up to get the full information Hamiltonian.
\begin{align}
\mathcal{H}(s,d) =&-\frac{1}{2} \mathrm{ln}\vert 2 \pi N \vert- \frac{1}{2} \mathrm{ln}\vert 2 \pi S \vert \nonumber\\
& +  \frac{1}{2}(d - Rs)^\dagger N^{-1} (d-Rs) \\ 
 &+ \frac{1}{2}s^\dagger S^{-1}s
\end{align}
With a quadratic completion the expression above can be again expressed as a Gaussian exponent:
\begin{align}
\mathcal{H}(s, d) \: \widehat{=}& \: \frac{1}{2} \left(s-m\right)^\dagger D^{-1} \left(s-m\right)\text{, with} \\
\label{eq:wf_cov}
D =& \:\left(R^\dagger N^{-1} R + S^{-1} \right)^{-1}\\
\label{eq:wf_j}
j =& \:R^\dagger N^{-1} d\\
\label{eq:wf_mean}
m =& \:Dj
\end{align}
Here, signal independent constants are dropped (as indicated by $\widehat{=}$), as the posterior distribution is obtained by normalizing the exponential of the negative Hamiltonian above over all possible signal configurations. Any signal independent constants do not contribute to the integral and are canceled out. In this case the Hamiltonian is of quadratic form with respect to the signal and we therefore just have to solve a Gaussian integral. The normalization turns out to be $\vert 2\pi D \vert ^{\frac{1}{2}}$. The posterior distribution of the signal $s$ given the data is also a Gaussian of the form
\begin{align}
\mathcal{P}(s\vert d,N,S) = \mathcal{G}(s - m,D) \text{.}
\end{align}
The posterior mean $m$, as given in Eq.\ \ref{eq:wf_mean}, is also called the Wiener filter solution, $j$ is called the information source (Eq.\ \ref{eq:wf_j}) and $D$ is the Wiener covariance (Eq.\ \ref{eq:wf_cov}).

\section{Critical Filtering}
\label{sec:CriticalFilter}

In many real world applications the prior correlation structure of the signal is unknown but often also of interest itself. Using the Bayesian framework we can build a hierarchical prior model which allows us to infer the correlation structure from the same data as the signal field itself. To do this we have to discuss a reasonable description of the correlation structure. A priori, no location is singled out and this should be reflected by the prior correlation structure. Initially all positions are equal, which corresponds to statistical homogeneity. Data, however, can break this symmetry, allowing for rich posterior correlations. With this assumption the prior correlation structure is diagonal in the harmonic domain of the signal field, according to the Wiener-Khintchin theorem \citep{Wiener, Khintchin}. For flat spaces this corresponds to the Fourier basis. This allows us to directly access the eigenbasis of the correlation structure, in which the operator is diagonal. We will use this property extensively in the derivation of the algorithm, as well as in its numerical implementations. With this, even high dimensional and high resolution applications are feasible. In the eigenbasis the prior correlation structure is completely described by its diagonal with the dimension of the original field, compared to the squared dimension of the operator structure in other functional bases. 

We can decrease the dimensionality of the correlation even further by assuming prior isotropy, which means to consider all directions to be statistically equal. This makes the correlation only dependent on the relative distance, not on its direction, collapsing the correlation structure to a one dimensional power spectrum function $p = p(\kappa)$ of the harmonic mode $\kappa$. All positions in the harmonic domain with the same distance from the center exhibit the same value. The relation between power spectrum $p$ and the diagonal correlation operator $S$ in the harmonic bases can be established using the power projection operator $\mathbb{P}$ and the harmonic transformation $\mathbb{F}$, which for flat spatial geometries corresponds to the Fourier transformation.

The power projection operator averages all positions corresponding to one spectral bin. A spectral bin is a subset of the whole harmonic domain $k \subset K$. The power projection $ \mathbb{P}$ integrates a harmonic field $a_\kappa$ according to each of those bins with $\kappa \in k$. This describes a new field $b$ with entries $b_k$ at the corresponding position. 

\begin{align}
b_k =\mathbb{P}_{k \kappa} a_\kappa =\frac{1}{\varrho_k}  \int_{\kappa \in k} a_\kappa 
\end{align}
Dividing by the bin volume $\varrho_k$ averages the field values over the bin.
By choosing rotationally symmetric bins we enforce isotropy. This allows us to describe the full correlation structure in terms of a power spectrum. We obtain the diagonal of the correlation operator $S$ in its harmonic domain by applying the adjoint power projection operator $\mathbb{P}^\dagger$ to a given power spectrum $p$. We indicate the transformation of a field to a diagonal operator in the fields domain by the $\widehat{\quad }$ symbol. The result of this operation is an operator with the property that all diagonal entries corresponding to one bin share the same value, given by the power spectrum. Applying harmonic transformations on both sides transforms the operator back to position space, as in the Wiener Filter example from the previous section. 
\begin{align}
S = \mathbb{F}^\dagger\left(\widehat{\mathbb{P}^\dagger p}\right)\mathbb{F}
\end{align}
It is worth pointing out that we obtain a proper projection operator by combining the power projection and its adjoint. Applying $\mathbb{P}^\dagger \mathbb{P}$ multiple times has no further effect, as after one application all values are averaged according to its binning. All underlying degrees of freedom of this projection are defined in the target of the power projector $\mathbb{P}$. This is the reason why it is convenient to parameterize the problem in this intermediate space.

In our application we will not only enforce radial symmetry but also perform a binning along the radial direction to reduce the number of spectral bins even further. The concrete choice depends on the problem. If a high spectral resolution is required, one does want to resolve the full spectral range. For other problems logarithmic binning with only a few bins might be sufficient. For each bin we will have to find a value $p_k$ which expresses its power. These are the parameters we characterize the correlation structure with. 

A fundamental property of the correlation structure is its positive definiteness, allowing only for positive eigenvalues. We will enforce this property by parametrizing $p$ on a logarithmic scale through a parameter $\tau$.
\begin{align}
p \equiv e^{\tau}
\end{align}
This also shifts the spectrum into the range of numerically convenient values, as physical fields towards smaller modes tend to decay on a logarithmic scale.
Returning to our Wiener filter problem, these considerations enter the signal prior, which now depends on the logarithmic power spectrum $\tau$.
\begin{align}
\mathcal{H}(s\vert \tau) = \frac{1}{2}\varrho ^\dagger \tau + \frac{1}{2} s^\dagger  \mathbb{F}^\dagger\left( \widehat{\mathbb{P}^\dagger e^{-\tau}}\right)  \mathbb{F} s
\end{align}
Here,  $\varrho$ is the volume of the individual bins in the harmonic space. The first term originates from the normalization of the prior $\mathrm{ln}\vert 2\pi S \vert$.
To apply Bayes' theorem we have to introduce distributions which capture our prior knowledge on the correlation structure. Some of them are already implicitly included in the model's construction. One additional assumption we often want to make is that the power spectrum is smooth on logarithmic scales. Some of this is also captured by the choice of binning. Large bins as well prohibit large variations in neighboring modes. A more elegant way to construct a smoothness prior is to measure deviations from a smooth spectrum in form of the second derivative $\Delta = \frac{\partial^2}{\partial y^2}$  on the logarithmic scale $y=\mathrm{ln} \kappa$ by taking its $L_2$ norm and weight it by some expected or tolerated deviation $\sigma$.  We can then write the prior spectral Hamiltonian similar to a Gaussian:
\begin{align}
\mathcal{H}(\tau) = \frac{1}{2 \sigma^2} \tau^\dagger \Delta^\dagger \Delta \tau
\end{align}
The derivative on a logarithmic scale exhibits some subtleties. The bin corresponding to the zero mode is now infinitely far away from all other modes, and therefore does not contribute to the expression above. On the boundary of the computational domain of the spectrum the curvature should vanish as well, corresponding to free boundary conditions. Therefore, the smoothness operator $\Delta^\dagger \Delta$, which measures how smooth its input is, is only positive semidefinite as it possesses vanishing eigenvalues and therefore cannot serve as a real covariance of a Gaussian. This makes the expression above an improper prior as it is not normalizable.

However, incorporating this prior into the overall model allows us to obtain valid posterior distributions. This is done by adding the smoothness Hamiltonian to the information Hamiltonian from above to get the full description of linear measurement with unknown correlation structure. 
This then reads:
\begin{align}
\label{eq:cf_ham}
\mathcal{H}(s,\tau\vert d) =& \frac{1}{2}(d - Rs)^\dagger N^{-1} (d-Rs) \nonumber \\ 
 &-  \frac{1}{2} \varrho^\dagger \tau + \frac{1}{2} s^\dagger  \mathbb{F}^\dagger\left( \widehat{\mathbb{P}^\dagger e^{-\tau}}\right)  \mathbb{F}s \nonumber \\
 & + \frac{1}{2 \sigma^2} \tau^\dagger \Delta^\dagger \Delta \tau
\end{align}
By this, the problem became highly nonlinear as we introduced coupling on all scales between the signal field and its prior correlation. We cannot calculate the posterior distribution analytically as it involves the normalization of the distribution with the Hamiltonian above with respect to both, the signal and its logarithmic power spectrum, which is unfeasible.
Therefore, an approximate approach to this problem must be taken. The most simple way is to maximize the Hamiltonian with respect to its parameters, leading to the joint maximum posterior solution (MAP).
Setting the derivative of the Hamiltonian with respect to the signal $s$ to zero while keeping the logarithmic power spectrum $\tau$ constant we recover the Wiener filter which minimizes this functional for a given power spectrum:
\begin{align}
D &= \left(R^\dagger N^{-1} R +  \mathbb{F}^\dagger\left( \widehat{\mathbb{P}^\dagger e^{-\tau}}\right)  \mathbb{F}  \right)^{-1} \label{eq:WF_correlation}\\
j &= R^\dagger N^{-1} d\\
m &= Dj \label{eq:WF_solution}
\end{align}
In this case $m$ corresponds to the current MAP solution. When setting the derivative of the Hamiltonian with respect to the logarithmic power spectrum to zero, the resulting equation cannot be solved analytically. Hence, one has to use other minimization schemes, preferably gradient based or fixed-point methods. The gradient reads
\begin{align}
\label{eq:MAPtaugrad}
\left.\frac{\delta \mathcal{H}(s,\tau\vert d)}{\delta \tau}\right\vert_{s=m} = -\frac{1}{2} m^\dagger \mathbb{F}^\dagger\widehat{\mathbb{P}^\dagger e^{-\tau}}\mathbb{F} m + \frac{1}{2 \sigma^2} \Delta^\dagger \Delta \tau \text{.}
\end{align}
Note that the hat in the equation above indicates that the expression below it is raised to a diagonal operator.
Once the minimum is reached we can use it to re-estimate the signal field and iterate this procedure until global convergence is achieved. 
The main flaw of this approach is that it does not pick up modes with low signal-to-noise ratio and therefore is only valid in the low noise regime. The Wiener filter suppresses such modes and consequently the power spectrum is strongly underestimated. The joint MAP estimate of signal and its spectrum exhibits a perception threshold \citep{ensslinfrommert} at moderate signal to noise ratios.

We can correct for this by informing the approximation about the uncertain nature of the Wiener filter reconstruction. There are multiple approaches which lead to the same correction. One way is to perform a renormalization calculation considering multiple loop diagrams in a field theoretical setting \citep{ensslinfrommert}. Another way is to replace the point estimate from the MAP approach by a Gaussian distribution which parameters are estimated using the Kullback-Leibler divergence \citep{ensslinweig}. Using these approaches, in the linear case the formulas for the signal field reconstruction remain unchanged, however a correction for the logarithmic power spectrum emerges. The condition becomes
\begin{align}
\label{eq:CF_solution}
0  \overset{!}{=} -\frac{1}{2} \mathrm{Tr}\left[\mathbb{P}\mathbb{F}\left(m m^\dagger + D\right)\mathbb{F}^\dagger\right]\widehat{ e^{-\tau}} + \frac{1}{2 \sigma^2} \Delta^\dagger \Delta \tau \text{,}
\end{align}
which can be solved for $\tau$ by (regularized) iterations.
By then alternatingly solving Eq.\ \ref{eq:WF_solution} and  Eq.\ \ref{eq:CF_solution} one obtains accurate estimates for the power spectrum even in high noise environments. Here, compared to \ref{eq:MAPtaugrad} the posterior Wiener covariance operator $D$ enters, correcting for uncertainty. This procedure is called the Critical Filter because it corresponds to a critical point in a phase diagram of filters of similar functional form \citep{ensslinfrommert}. 

\section{Posterior sampling}
\label{sec:sampling}
To estimate the power spectrum we have to minimize the  KL-divergence. We can only access it stochastically by using posterior samples of the excitation field. Obtaining these samples requires some effort, but we can exploit the Wiener filter structure of the approximate excitation posterior to draw independent samples. The procedure is analogous to the one discussed in \citep{NICAAC}, however, for completeness we will briefly outline it here as well. The idea is that we set up a mock observation of a synthetic signal drawn from the prior, which is then reconstructed using the Wiener filter. The residual between the mock reconstruction and the initial signal exhibits the correct correlation structure and correcting for the true mean yields a sample from the approximate posterior, which we can then use to estimate the KL.

We start with a sample $\xi'$ from the excitation prior, i.e. a white Gaussian field in the harmonic domain.
\begin{align}
\xi' \curvearrowleft\mathcal{G}(\xi',\mathbb{1})
\end{align}
The next step is to identify the effective instrument response to recover the correct Wiener filter covariance from Eq.\ (\ref{eq:approx_WF_cov}). This linearized response reads
\begin{align}
R^* = Rf'(At)A \text{.}
\end{align}
To set up a synthetic measurement we also need  a new synthetic noise contribution $n'$, which we draw from the noise prior
\begin{align}
n' \curvearrowleft \mathcal{G}(n',N) \text{.}
\end{align}
With all those components combined we can imitate a linear measurement to get mock data. This is exactly the situation we discussed at the beginning of section \ref{sec:WienerFilter}.
\begin{align}
d' = R^* \xi' + n'
\end{align}
Based on this virtual measurement we know how to reconstruct the synthetic excitation field. Using the Wiener filter equations its posterior mean is
\begin{align}
\label{eq:sampleWF}
t' = \Xi j' \:\text{, with}\\
j' = R^{*\dagger} N^{-1}d'  \text{.}
\end{align}
Here $\Xi$ corresponds to the curvature of the problem Hamiltonian with respect to the excitation, as it was defined in \ref{eq:approx_WF_cov}.
Now we can verify that the residual $ \xi' - t'$ exhibits the correct internal correlation structure of the approximate excitation posterior.
\begin{align}
\langle (\xi'-t')(\xi'-t')^\dagger \rangle = \Xi
\end{align}
For the posterior samples we demand the same correlation structure, therefore we write:
\begin{align}
\xi^* - t  =: \xi' -t' \text{.}
\end{align}
Solving this equation for the sample $\xi^*$ provides us with the posterior sample. It corrects the mock residual by the correct mean.
\begin{align}
\xi^* = \xi' - t' + t
\end{align}
These samples exhibit the desired correlation structure and mean. Hence, they are samples from the approximate posterior distribution.
\begin{align}
\xi^* \curvearrowleft \mathcal{G}(\xi^* - t, \Xi)
\end{align}
For the estimation of the power spectrum we might require multiple samples to reduce sampling noise of the KL estimate. Those are obtained by repeating the described procedure with different prior  and noise samples. However, we might want to reduce the number of samples as far as possible as each of them requires solving  a Wiener filter problem (Eq. \ref{eq:sampleWF}), which is numerically roughly as complex as the excitation estimate. We propose to start with a few samples and then increase their number during the inference for ultimately high accuracy results.

\end{document}